\documentclass[12pt, draftclsnofoot, onecolumn]{IEEEtran}
\usepackage{graphicx}
\usepackage{epsfig}
\usepackage{algorithm}
\usepackage{algorithmic}
\usepackage{ifmtarg}
\usepackage{cite}
\usepackage{amssymb}
\usepackage{amsthm}
\usepackage[fleqn]{amsmath}
\usepackage{amsmath}
\usepackage{color}
\usepackage{bbm}
\usepackage{cite}
\usepackage{epstopdf}
\usepackage[table,xcdraw]{xcolor}
\usepackage{flushend}
\newtheorem{theorem}{\textbf{\text{Theorem}}}
\newtheorem{lemma}{\textbf{\text{Lemma}}}
\newtheorem{proposition}{Proposition}

\newtheorem{assumption}{Assumption}
\newtheorem{remark}{Remark}
\usepackage{flushend}
\usepackage{subcaption}
\usepackage{tikz,pgfplots}
\usetikzlibrary{plotmarks}
\newlength\figureheight
\newlength\figurewidth

\begin{document}
\title{In-Band $\alpha$-Duplex Scheme for Cellular Networks: A Stochastic Geometry Approach}
\author{
\IEEEauthorblockN{\large  Ahmad AlAmmouri, Hesham ElSawy, Osama Amin, and Mohamed-Slim Alouini}

\thanks{The authors are with Computer, Electrical, and Mathematical Sciences and Engineering (CEMSE) Divison, King Abdullah University of Science and Technology (KAUST), Thuwal, Makkah Province, Saudi Arabia. (Email: \{ahmad.alammouri, hesham.elsawy, osama.amin, slim.alouini\}@kaust.edu.sa) }}

\maketitle
\thispagestyle{empty}
\pagestyle{empty}

\begin{abstract}
In-band full-duplex (FD) communications have been optimistically promoted to improve the spectrum utilization and efficiency. However, the penetration of FD communications to the cellular networks domain is challenging due to the imposed uplink/downlink interference. This paper presents a tractable framework, based on stochastic geometry, to study FD communications in cellular networks. Particularly, we assess the FD communications effect on the network performance and quantify the associated gains. The study proves the vulnerability of the uplink to the downlink interference and shows that FD rate gains harvested in the downlink (up to $97\%$) come at the expense of a significant degradation in the uplink rate  (up to $94\%$). Therefore, we propose a novel fine-grained duplexing scheme, denoted as $\alpha$-duplex scheme, which allows a partial overlap between the uplink and the downlink frequency bands. We  derive the required conditions to harvest rate gains from the $\alpha$-duplex scheme and show its superiority to both the FD and half-duplex (HD) schemes. In particular, we show that the  $\alpha$-duplex scheme provides a simultaneous improvement of $28\%$ for the downlink rate and $56\%$ for the uplink rate. Finally, we show that the amount of the overlap can be optimized based on the network design objective. 
\end{abstract}


\section{Introduction}\label{Introduction}
    
Due to the overwhelming effect of self-interference (SI), wireless transmission and reception are always separated in time, denoted as time division duplexing (TDD), or in frequency, denoted as frequency division duplexing (FDD). Recent advances in transceiver design tend to make SI cancellation (SIC) viable and alleviate the necessity of such time/frequency separation \cite{InBand2014Sabharwal, Applications2014Hong,Prototyping2015Chung}. That is; SIC techniques enable transceivers to achieve acceptable isolation between transmit and receive circuitries while transmitting and receiving on the same time-frequency resource block, either by using single \cite{Full2013Bharadia} or multiple \cite{Design2014Duarte} antennas. It is argued that exploiting the entire bandwidth (time) for FDD (TDD) systems for transmission and reception, denoted as in-band FD communications, can double the spectral efficiency and improve the network capacity\cite{InBand2014Sabharwal, Full2013Bharadia}. This argument makes the in-band FD scheme  a good candidate technology for cellular operators to cope with the challenging performance metrics defined for 5G cellular networks \cite{Applications2014Hong,A2015Kim}.

In the context of cellular networks, FD communications directly imply simultaneous uplink (UL)/downlink (DL) transmissions on the same time/frequency resource blocks. Note that the SIC techniques enable transceivers to cancel only their SI, but not the interference originated from other sources reusing the same frequency over the spatial domain. Hence, in a large scale cellular network with spatial frequency reuse, FD communications impose inter-cell UL/DL interference, hereafter denoted as cross-mode interference. Since cellular networks are recognized as interference limited networks, the cross-mode interference may diminish the harvested performance gains \cite{Full2015Goyal}.  Hence, an explicit study of the UL and DL performances under cross-mode interference is essential to characterize the FD performance and quantify the associated gains. In this regard, stochastic geometry can be exploited to model the interference and achieve such performance characterization\cite{Stochastic2013ElSawy, Stochastic2012Haenggi,Modeling2016ElSawy}.

Several research efforts are exerted to study the effect of cross-mode interference on the  FD performance for different types of large-scale wireless networks. In the context of ad-hoc networks,  authors in \cite{Throughput2015Tong2,Does2015Xie,A2015Tong} show that FD communications can improve the overall throughput despite the increased aggregate interference level. In the context of cellular networks, FD communications' gains are mainly quantified for the DL performance. Assuming perfect SIC, the results in \cite{On2014Alves} show that FD communications almost double the spectral efficiency for the DL. Even with imperfect SIC, it is shown in \cite{Hybrid2015Lee,Full2015Mohammadi,Full2015Atzeni} that FD communications can improve the DL spectral efficiency. However, the models in \cite{On2014Alves, Hybrid2015Lee, Full2015Mohammadi,Full2015Atzeni} overlook the effect of FD communications on the UL performance.  Note that the effect of the cross-mode interference is more prominent on the UL due to the high disparity between the UL and DL transmissions (e.g., power level, interference protection, etc.). In particular, due to the high transmission power of the base stations (BSs) along with the vulnerability of the UL transmission \cite{Load2014AlAmmouri,On2014ElSawy, Analytical2013Novlan}, the cross-mode interference on the UL performance is the bottleneck of the FD operation. Assuming perfect SIC, the impact of cross-mode interference on the UL performance is highlighted in \cite{Analyzing2013Goyal} by showing that the FD gains are mainly from the DL direction. Hence, the authors in \cite{Analyzing2013Goyal} proposed a scheduling algorithm to improve the UL performance under FD operation.\footnote{The authors in \cite{Throughput2015Mahmood} highlights that the asymmetric nature of practical UL/DL transmissions is another challenge in FD communications.}

This paper presents a mathematical framework, based on stochastic geometry, for FD communications in cellular networks\footnote{This work is an extension of \cite{InBand2015Ahmad}, in which we extend the analysis to account for the explicit performance of cell center and cell edge users. We also analyzed effective rate, ergodic rate, and the outage probability in addition to the bit error probability (BEP) analyzed in \cite{InBand2015Ahmad}. Last but not least, more numerical results and insights are presented.}. The developed framework accounts for the explicit UL and DL performances. It also captures the impact of realistic network parameters such as  pulse-shaping, matched filtering, per user UL power control, and limited transmission power of users equipment (UEs). The obtained results are consistent with the literature and confirm the positive impact of FD communications on the DL performance. However, to the best of our knowledge, this paper is the first to prove the vulnerability of the UL to the cross-mode interference. In particular, we show that the DL performance enhancement by FD communications (up to $97\%$)  may come at the expense of severe degradation in the UL performance (up to $94\%$). Therefore, we propose a novel fine-grained duplexing scheme, denoted as $\alpha$-duplex, which allows partial overlap between the UL and the DL frequency bands\footnote{There are schemes that propose partially overlapping adjacent channels to trade SINR for BW in half-duplex systems\cite{On2013Hamdi, Multiple2009Hoque, Efficient2011Abbasi}. However, to the best of our knowledge, partial overlap with pulse-shaping is not studied for FD systems.}. The amount of the overlap is controlled via the design parameter $\alpha$ to balance the trade-off between the UL and DL performances. It is worth mentioning that the proposed $\alpha$-duplex scheme captures the FD and traditional half-duplex (HD) systems as special cases. Specifically, setting $\alpha$ to one enforces FD communications while setting $\alpha$ to zero maintains HD communications. To this end, we show that the  $\alpha$-duplex scheme provides a simultaneous improvement of $28\%$ for the DL rate and $56\%$ for the UL rate. Finally, we show that the amount of the overlap can be optimized based on the network design objective. 

The rest of the paper is organized as follows: Section II, presents the system model and assumptions. Section III, analyzes the performance of the $\alpha$-duplex system. Numerical \& simulation results along with remarks on the $\alpha$-duplex system design are presented in Section IV before presenting the conclusion in Section V.

\textit{\textbf{Notations}}: $\mathbb{E} [\cdot]$ denotes the expectation over all the random variables (RVs) inside $[\cdot]$, $\mathbb{E}_{x} [\cdot]$ is the expectation with respect to (w.r.t.) the RV $x$, $\mathbbm{1}_{\{\cdot\}}$ is the indicator function which takes the value $1$ if the statement $\{\cdot\}$ is true and takes the value $0$ otherwise, $.*$ is the convolution operator and $S^{*}$ is the complex conjugate of $S$.

\section{System Model}\label{System Model}
\subsection{Network Model}
We consider a bi-dimensional  single-tier cellular network where the locations of the BSs\footnote{We assume that the BSs are equipped with a single antenna, extending the results to capture multi-input-multi-output (MIMO) systems with FD communications can be done following \cite{Directional2016Psomas,Full2015Atzeni}.} are modeled as a homogeneous Poisson point process (PPP) ${\Psi}_{\rm d}=\{x_i,i=1,2,3,.... \}$ with intensity $\lambda$, where $x_i \in \mathbb{R}^2$ denotes the location of the $i^{\rm th}$ BS. Besides simplifying the analysis, the PPP assumption for cellular networks is validated by experimental and theoretical studies \cite{A2011Andrews, Spatial2013Guo,Stochastic2015Lu, Using2013Blaszczyszyn} and is currently use to model FD communications in cellular networks \cite{On2014Alves, Hybrid2015Lee, Full2015Mohammadi,Full2015Atzeni,Analyzing2013Goyal}. The locations of the UEs are modeled via an independent PPP  $\Psi_{\rm{u}}$ with intensity $\lambda_{\rm{u}}$, where $\lambda_{\rm{u}}>>\lambda$ such that each BS has at least one UE within its association area. Radio signal strength based association is adopted, which boils down to the nearest BS association in the depicted single-tiered network. A general power-law path-loss model is assumed in which the signal power decays at the rate $r^{-\eta}$ with the distance $r$, where $\eta>2$ is the path-loss exponent \cite{Wireless2005Goldsmith}. In addition to the path-loss attenuation, UL and DL signals experience  Rayleigh fading with independent and identically distributed  (i.i.d) channel gains, and hence, the channel power gains are exponentially distributed random variables. Without loss of generality, we assume that all channel gains have unit mean.\footnote{It is worth mentioning that the Rayleigh fading assumption is selected to simplify the analysis and expressions, the proposed framework can be extended to capture different fading channels as in \cite{Average2013Renzo,Modeling2016AlAmmouri}.}.

All BSs transmit at a constant power level of $P_{\rm{d}}$ in the DL. In the UL, UEs employ channel inversion power control scheme with target power level $\rho$ and maximum transmit power constraint of  $P^{({\rm{M}})}_{\rm{u}}$. The channel inversion power control is a special case of the fractional power control policy recommended by the 3GPP~\cite{3GPP_Power}, which aims at maintaining a unified target average power level of $\rho$ at the serving BSs. Due to the irregular structure of the Poisson-Voronoi tessellation, UEs are classified into cell edge users (CEUs) and cell center users (CCUs) according to their path-losses rather than their actual location w.r.t. the cell edge. Specifically, CCUs are those who can invert their path-loss to achieve the required power level of $\rho$ at their serving BSs. On the other hand, CEUs are those who experience high path-losses, and hence, transmit at their maximum power yet cannot achieve the required power level of $\rho$ at their serving BSs. Without loss of generality, we assume a single pair of channels  (i.e., one for the UL and the other for the DL) which is universally reused across the network with no intra-cell interference. Extension to the multi-channel case with load-awareness is straightforward by following the methodology in \cite{Load2014AlAmmouri}.
 
\normalsize
\subsection{$\alpha$-Duplex Model}

\begin{figure}[t]
\centerline{\includegraphics[width=  6 in]{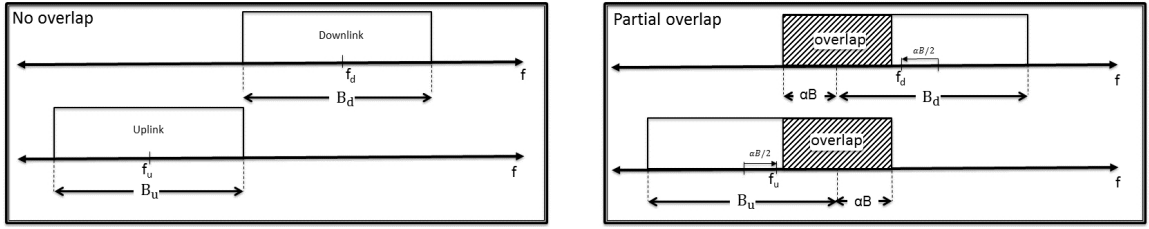}}\caption{\,A schematic diagram of the proposed $\alpha$-duplex scheme.}
\label{fig:proposed}
\vspace{-0.8cm}
\end{figure}

Both the BSs and UEs have SIC capabilities and can operate in the FD mode.  We focus on the case in which the DL and UL channels occupy two non-overlapping adjacent null-to-null bands\footnote{The analysis can be directly extended to non-adjacent bands via carrier aggregation. Also, the effect of guard bands can be easily incorporated into the analysis.}. The DL and UL bandwidths (BWs) are denoted by $B_{\rm{d}} $ and $B_{\rm{u}} $\footnote{Although symmetric traffic applications, such as social networking, video calls, real-time video gaming, etc. requires symmetric UL/DL rates \cite{Analytical2013Novlan}, we can set the $B_{\rm{d}} $  and $B_{\rm{u}} $ in order to capture possible UL/DL traffic asymmetry. }, respectively, in which the carrier frequencies $f_{\rm{d}} > f_{\rm{u}}$. SIC capability is exploited to increase the spectral efficiency and allow partial overlap of $2 \alpha B$, where $B = \min(B_{\rm{u}} ,B_{\rm{d}} )$, between the UL and the DL transmissions, as shown in Fig.~\ref{fig:proposed}. Note that $\alpha$ is a system level parameter used by all BSs and all UEs. The $\alpha$-duplex scheme extends the UL channel BW from $B_{\rm{u}} $ to $B_{\rm{u}}  + \alpha B$, where $\alpha B$ is consumed from the adjacent DL band. Similarly, the DL channel BW is extended from $B_{\rm{d}}$ to $B_{\rm{d}}  + \alpha B$, where $\alpha B$ is consumed from the adjacent UL band. It is worth noting that, since the transmission BW is a function of $\alpha$, the center frequencies for the UL and DL  frequency bands are also functions of $\alpha$. According to our system model, the difference between the DL center frequency ($f_{\rm{d}} $) and the UL center frequency ($f_{\rm{u}} $), denoted as   $\Delta f(\alpha)$,  is given by,
\small
\begin{align}
   \Delta f(\alpha) =  f_{\rm{d}} -f_{\rm{u}} =\frac{B_{\rm{u}} +B_{\rm{d}} }{2} - \alpha B.
\end{align}

\normalsize
As a result of the UL/DL spectrum overlap, the test receiver will experience interference from DL BSs as well as UL UEs. 

\subsection{base-band Signal Representation}

Hereafter, for notational convenience and to avoid repetitions, we use the notations ${ \chi} \in \{\rm{u},\rm{d}\}$,  $\bar{\rm \chi} \in \{\rm{u},\rm{d}\}$, where ${ \chi} \neq \bar{ \chi}$, to express generic formulas that hold true for the UL and the DL transmissions. Hence, whenever applicable, a single expression and/or discussion is valid for both transmissions.

The data at the test transmitter of the system ${\chi}$, which is a UE in the UL or a BS in the DL, is mapped to a bi-dimensional and symmetric constellation with unit energy. All transmitters from the same system ${\chi}$ use a unified pulse shape $s_{\chi} (t) \overset{FT}{\longleftrightarrow} S_{\chi} (f)$, where $FT$ denotes the Fourier transform. At the receiver side, the base-band signal (i.e., after down conversion) is passed through a matched and low pass filters $H_{\chi}(f)$ before sampling at the input of the decoder. The combined frequency domain representations for the matched and low pass filters in the UL and the DL are given by,
\small
\begin{align}
H_{\chi}(f)=\left\{
	\begin{array}{ll}
		S_{\chi} ^{*}(f) \ \ \ \ \    -\frac{B_{\chi}+\alpha B}{2} &\leq f \leq \frac{B_{\chi}+\alpha B}{2}, \\
		0 & \! \!\! \!\! \! \rm{elsewhere},
	\end{array}
\right.
\end{align}

\normalsize
At the test receiver side (BS for the UL or UE for the DL), the received base-band signal at the input of the matched filter can be expressed~as
\small
\begin{align}
\!\!\!\!\!\!\!\!\!\! y_{\chi}(t)=  A \sqrt{P_{r_o} h_o} s_{\chi}(t) + \sum_{k \in \tilde{\Psi}_{\chi}} i_{k}^{(\chi)}(t)+ \sum_{j \in \tilde{\Psi}_{\bar{\chi}}}i_{j}^{(\bar{\chi})}(t) + i_{\rm s}(t) + n_{o}(t).
\label{base_band1}
\end{align}
\normalsize
where $A$ represents the complex symbol of interest, $P_{r_o}$ is the average received power at the test receiver, $h_o$ is the intended channel fading power gain, $\tilde{\Psi}_{\chi} \subset \Psi_{\chi}$ is the set of intra-mode interferers, $ \sum_{k \in \Psi_{\chi}} i_{k}^{(\chi)}(t)$  is the aggregate intra-mode interference, $\tilde{\Psi}_{\bar{\chi}} \subset \Psi_{\bar{\chi}}$ is the set of cross-mode interferers, $\sum_{j \in \Psi_{\bar{\chi}}} i_{j}^{(\bar{\chi})}(t) $ is the aggregate cross-mode interference,  $i_{\rm s}(t)$  is the residual self-interference (i.e., after digital and analogue cancellation)\footnote{In this work, we assume that both the BSs and the UEs have SI cancellation capabilities and operate in FD mode, for FD BSs with traditional HD UEs refer to  \cite{Flexible2016AlAmmouri}.}, and $n_o(t)$ is a white complex Gaussian noise with zero mean and two-sided power spectral density $N_o/2$. 

To facilitate the analysis, we abstract symbols from interfering sources via Gaussian codebooks as in \cite{A2009Shobowale, Error2015Afify, Performance2015Afify}. The accuracy of the Gaussian codebook approximation for interfering symbols from several constellation types have been verified in \cite{Influence2005Giorgetti, The2015Afify}. In this case, the intra-mode, cross-mode, and self-interference terms can be expressed, respectively, as
\small
\begin{align} \label{inter1}
&i_{k}^{( \chi)}(t)=  \zeta_{\chi_k} s_{\chi}(t) \sqrt{P_{\chi_k} h_{\chi_k} r_{\chi_k}^{-\eta}}, \notag \\
&i_{j}^{(\bar{\chi})}(t)=  \zeta_{\bar{\chi}_j} s_{\bar{\chi}}(t) \sqrt{P_{{\bar{\chi}}_j} h_{\bar{\chi}_j} r_{\bar{\chi}_j}^{-\eta}} \exp \left( j 2 \pi\Delta f(\alpha) t \right),  \notag \\
&i_{s}(t)=  \zeta_{s} \sqrt{\beta P_{\bar{\chi}_o}} s_{\bar{\chi}}(t) \exp \left( j 2 \pi\Delta f(\alpha) t \right),
\end{align}\normalsize
where $\zeta_{\chi_k}$, $\zeta_{\bar{\chi}_j}$, and $\zeta_s$ are independent unit variance circularly symmetric complex Gaussian symbols. $h_{\chi_k}$'s and $h_{\bar{\chi}_j}$'s are the channel fading power gains, $r_{\chi_k}$ and $r_{\bar{\chi}_j}$ are the distances between the tagged receiver and the $k^{\rm th}$ intra-mode interferer and the $j^{\rm th}$ cross-mode interferer, respectively. $P_{\chi_k}$ is the transmitted power of the $k^{\rm th}$ intra-mode interferer and $P_{\bar{\chi}_j}$ is the transmitted power of the $j^{\rm th}$ cross-mode interferer. $\beta$ represents the self-interference attenuation, which is set to zero if perfect SI is achieved. It is worth mentioning that the self-interference is a special type of cross-mode interference at which the interferer and the test receiver are collocated. Hence, the phase-shift of $\exp \left( j 2 \pi\Delta f(\alpha) t \right)$ appears in the cross-mode interference and self-interference terms in \eqref{inter1} to capture the $\Delta f(\alpha)$ offset between the center frequencies of the UL and DL.

\section{Performance Analysis} 

This work explicitly characterizes the UL and the DL performances in terms of bit error probability (BEP), outage probability and transmission rate for the $\alpha$-duplex scheme. While the outage probability and the BEP are important key performance indicators, they are not sufficient to characterize the $\alpha$-duplex operation. This is because the outage probability and the BEP are independent of the BW and are only affected by the cross-mode interference. The picture is complete by looking into the transmission rate which captures the effect of both the cross-mode interference as well as the improved BW. For the sake of complete exposition, we consider both the ergodic rate for CSI aware systems as well as the throughput for fixed rate CSI unaware systems. To characterize the $\alpha$-duplex system, we start the analysis by looking into the effect of pulse-shaping, filtering, and duplexing on the SINR at the input of the decoder in Section~\ref{filtering}. We then characterize cross-mode and intra-mode interferences by the Laplace transform (LT) of their distributions in each network scenario (i.e., UL and DL) in Section~\ref{LTS}. Finally, the representation of each of the aforementioned performance metrics in terms of the LTs of interferences is presented in Sections~\ref{Perform1}, \ref{Perform2}, and \ref{Perform3}. Without loss in generality, the analysis are conducted for a test receiver (once for UL and once for DL) located at the origin.\footnote{The origin is an arbitrary reference point in the $\mathbb{R}^2$ plane that is usually selected at the test receiver to simplify notations.} According to Slivnyak's theorem \cite{Stochastic2012Haenggi}, any other location in the space has an identical statistical behavior to the origin.

\subsection{The Effect of Pulse-Shaping, $\alpha$-Duplexing, and Filtering} \label{filtering}

At the receiver side of $\chi$, the matched filter convolves the received base-band signal (i.e., down-converted) with the conjugated time-reversed pulse shape template  $s_\chi(t)$. Then, the output of the matched filter is fed to a low-pass filter with the BW of interest (i.e., $B_\chi+\alpha B$). The output of the low-pass filter is sampled at $t_o$ and fed to the ML decoder. From \eqref{base_band1}, the sampled base-band signal at the input of the decoder is given by,
\small
\begin{align}
\!\!\!\!\!\!\!\!\!\! y_{\chi}(t_o)&=  \left. \left(A \sqrt{P_{r_o} h_o} s_{\chi}(t) + \sum_{k \in \tilde{\Psi}_\chi} i_{k}^{(\chi)}(t)+ \sum_{j \in \tilde{\Psi}_{\bar{\chi}}}i_{j}^{(\bar{\chi})}(t) + i_{\rm s}(t) + n_{o}(t)\right).*h_\chi(t-t_o) \right|_{t=t_o}.
\label{base_band_h1}
\end{align}

\normalsize
Using the distributive property of the convolution operator and substituting \eqref{inter1} in \eqref{base_band_h1}, the signal in \eqref{base_band_h1} can be rewritten as:

\small
\begin{align}
\!\!\!\!\!\!\!\!\!\!\!\!\! y_{\chi}(t_o)=& A \sqrt{P_{r_o} h_o} \mathcal{I}_{\chi}(\alpha) + \sum_{k \in \tilde{\Psi}_{\chi}} \zeta_{\chi_{k}} \sqrt{P_{\chi_k} h_{\chi_{k}} r_{\chi_{k}}^{-\eta}}   \mathcal{I}_{{\chi}}(\alpha) + \sum_{j \in \tilde{\Psi}_{\bar{\chi}}}\zeta_{{\bar{\chi}_{j}}} \sqrt{P_{\bar{\chi}_j} h_{\bar{\chi}_{j}} r_{\bar{\chi}_{j}}^{-\eta}} \mathcal{C}_{\chi}(\alpha) +  \zeta_{{\rm{s}}} \sqrt{\beta P_{\chi_o}} \mathcal{C}_{\chi}(\alpha) + n_{\chi},
 \label{mathced_downlink1}
\end{align}
\normalsize
where $\mathcal{I}_{\chi}(\alpha)$ and $\mathcal{C}_{\chi}(\alpha)$ are the effective amplitude factors in, respectively, the intra- and cross-mode signals due to filtering and pulse-shaping. $\mathcal{I}_{\chi}(\alpha)$ and $\mathcal{C}_{\chi}(\alpha)$ are given by

\small
\begin{align}
\mathcal{I}_{\chi}(\alpha) \quad {=}  \left. s_{\chi}(t) .* h_{\chi}(t-t_o) \right|_{t=t_o}  \quad {=} \int\nolimits_{-\frac{B_{\chi}+\alpha B}{2}}^{\frac{B_{\chi}+\alpha B}{2}} S_{\chi}(f) S_{\chi}^{*}(f) df,
\label{equ:Idd1}
\end{align}
\normalsize
and
\small
\begin{align}
\mathcal{C}_{\chi}(\alpha) \quad  {=}   \left. s_{\bar{\chi}}(t) \exp \left( j 2 \pi\Delta f(\alpha) t \right) .* h_{\chi}(t-t_o) \right|_{t=t_o}   \quad   {=} \int\nolimits_{-\frac{B_{\bar{\chi}}+\alpha B}{2}}^{\frac{B_{\bar{\chi}}+\alpha B}{2}} S_{\bar{\chi}}(f- \Delta f (\alpha)) S_{\chi}^{*}(f) df,
\label{equ:Idd2}
\end{align}
\normalsize
where the second equalities in \eqref{equ:Idd1} and \eqref{equ:Idd2} follow from the frequency domain representation of the convolution and sampling processes. $n_\chi$ in \eqref{mathced_downlink1} is the AWGN in which pulse-shaping and filtering reduce the noise power at the decoder to
\small
\begin{align}
    \sigma_{n_{\chi}}^2=N_o \int\nolimits_{-\frac{B_{\chi}+\alpha B}{2}}^{\frac{B_{\chi}+\alpha B}{2}} |H_{\chi}(f)|^2 df=N_o |\mathcal{I}_{\chi}(\alpha)|^2.
\end{align}
\normalsize

While \eqref{mathced_downlink1}, \eqref{equ:Idd1}, and \eqref{equ:Idd2} show the effect of pulse-shaping and filtering on the received base-band signal, it is more important to see the pulse-shaping and filtering effect on the SINR. Inspecting \eqref{mathced_downlink1}, it is clear that the aggregate interference is Gaussian if we condition on the network geometry, channel gains, and transmission powers\footnote{The transmission power of the UEs are random due to the employed channel inversion power control.}. Hence, the aggregate interference and noise terms can be lumped together into a conditional Gaussian random variable with the total variance. Let $\Xi_{\chi}=\{h_o,P_{r_o}, P_{\chi_k} h_{\chi_{k}} , r_{\chi_{k}},P_{\bar{\chi}_j}, h_{\bar{\chi}_{j}},  r_{\bar{\chi}_{j}}\}$, then the conditional SINR can be expressed as

\small
{\begin{align} \label{SINR_chi}
  \!\!\!\!\!\!\!\!\!\!\!\!\! {\rm SINR}_{\chi} \left(\alpha \vert \Xi_{\chi} \right) &= \frac{| \mathbb{E} \left[ y_{\chi}(t_o)\big \vert  \Xi_{\chi} \right] |^2}{\text{Var}\left( y_{\chi}(t_o) \big \vert  \Xi_{\chi}  \right)}, \notag \\ 
 & = \frac{ P_{r_o} h_o |\mathcal{I}_{\chi}(\alpha)|^2}{\underset{{k \in {\Psi}_{\chi}}}{\sum}  P_{\chi_k} h_{\chi_{k}} r_{\chi_{k}}^{-\eta}|\mathcal{I}_{\chi}(\alpha)|^2 + \underset{j \in {\Psi_{\bar{\chi}}}}{\sum} P_{{\bar{\chi}}_j} h_{\bar{\chi}_{j}} r_{\bar{\chi}_{j}}^{-\eta}|\mathcal{C}_{\chi}(\alpha)|^2+\beta P_{\bar{\chi}_o} |\mathcal{C}_{\chi}(\alpha)|^2+N_o |\mathcal{I}_{\chi}(\alpha)|^2 } , \notag \\
& = \frac{ P_{r_o} h_o}{\underset{{k \in {\Psi}_{\chi}}}{\sum}  P_{\chi_k} h_{\chi_{k}} r_{\chi_{k}}^{-\eta} + \underset{j \in {\Psi_{\bar{\chi}}}}{\sum} P_{{\bar{\chi}}_j} h_{\bar{\chi}_{j}} r_{\bar{\chi}_{j}}^{-\eta}|\tilde{\mathcal{C}}_{\chi}(\alpha)|^2+\beta P_{\bar{\chi}_o} |\tilde{\mathcal{C}}_{\chi}(\alpha)|^2+N_o  } ,
\end{align}}\normalsize
where $\tilde{\mathcal{C}}_{\chi}(\alpha) = \frac{{\mathcal{C}}_{\chi}(\alpha)}{{\mathcal{I}}_{\chi}(\alpha)}$. Equation \eqref{SINR_chi} confines the effect of pulse-shaping and filtering to the cross-mode interference terms. As a result of pulse-shaping and filtering, only a fraction of  ($|\tilde{\mathcal{C}}_{\chi}(\alpha)|^2$) from the cross-mode interference power leaks to input of the decoder of $\chi$. Hereafter, ($|\tilde{\mathcal{C}}_{\chi}(\alpha)|^2$) is denoted as the effective cross-mode interference power factor, which measures the amount of cross-mode interference power within the BW of interest at the input of the decoder. Since the SI is a special type of cross-mode interference, pulse-shaping and filtering reduce the SI power with the factor  $|\tilde{\mathcal{C}}_{\chi}(\alpha)|^2$ in addition to the built-in SIC factor $\beta$. 

\normalsize
\begin{figure*}[t!]
    \centering
    \begin{subfigure}[t]{0.31\textwidth}
  \centerline{\includegraphics[width=  2.2in]{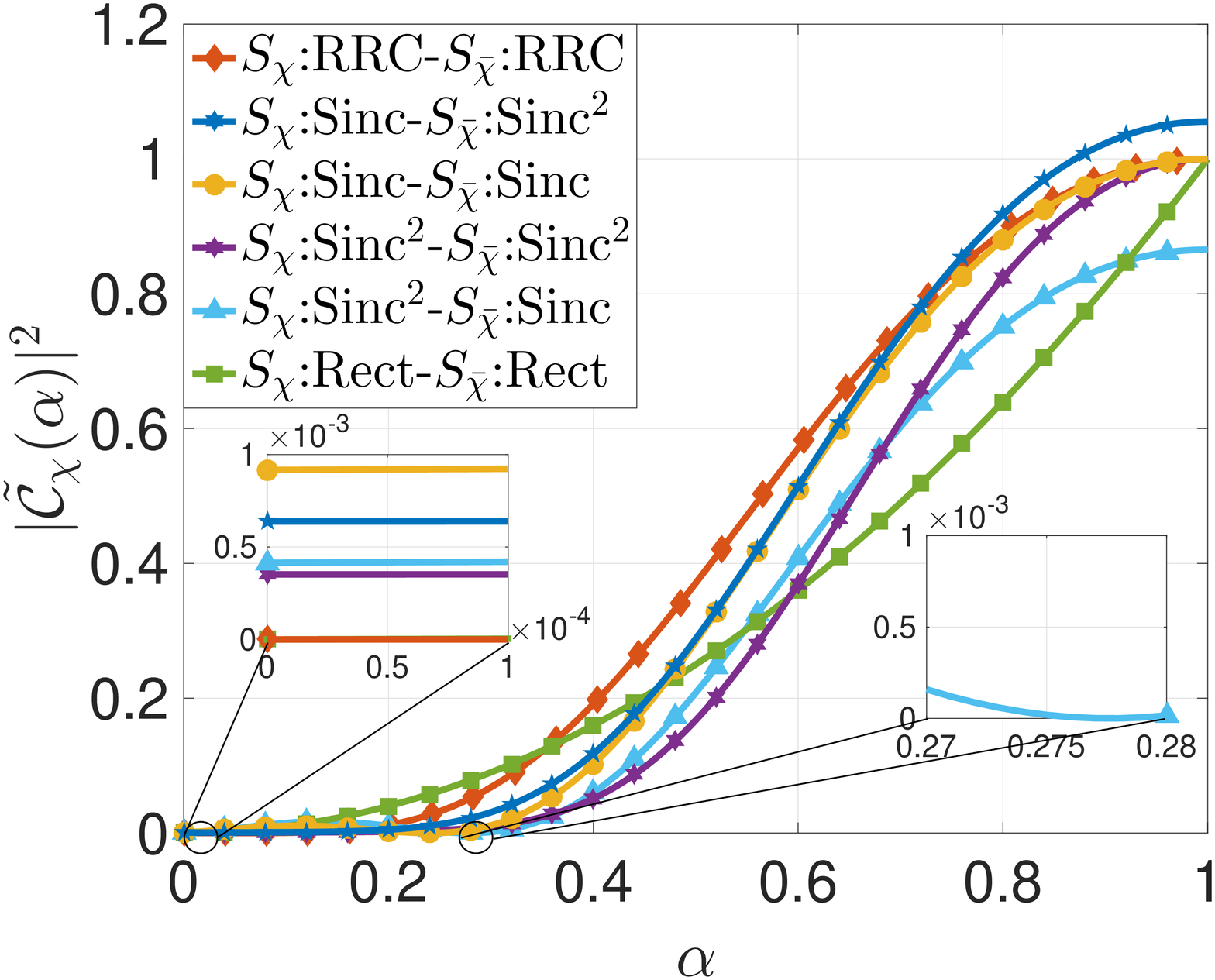}}
      \caption{\, $|\tilde{\mathcal{C}}_{\chi}|^2$ vs $\alpha$ for rectangular (Rect), root raised cosine (RRC), ${\rm Sinc}$ and ${\rm Sinc^2}$ pulse shapes.}
\label{fig:Effective}
    \end{subfigure}
    ~ 
    \begin{subfigure}[t]{0.31\textwidth}
       \centerline{\includegraphics[width=  2.2in]{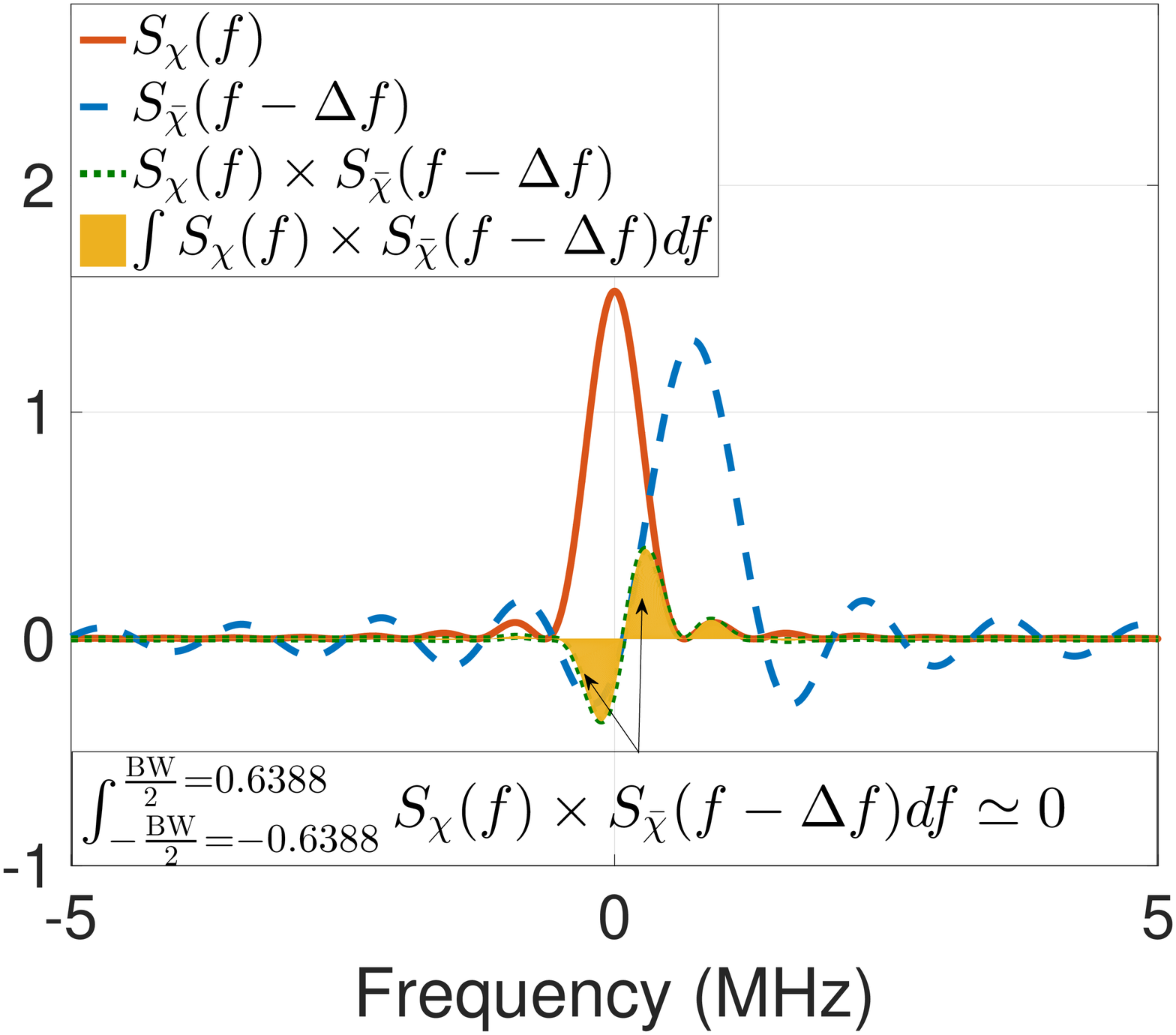}}\caption{\, Frequency domain illustration of the Sinc$^2$ ($S_{\chi}$) and Sinc ($S_{\bar{\chi}}$) pulse shapes at $\alpha_{\rm sp}=0.2776$.}
\label{fig:MinAlpha}
    \end{subfigure}
    ~
    \begin{subfigure}[t]{0.31\textwidth}
       \centerline{\includegraphics[width=  2.2in]{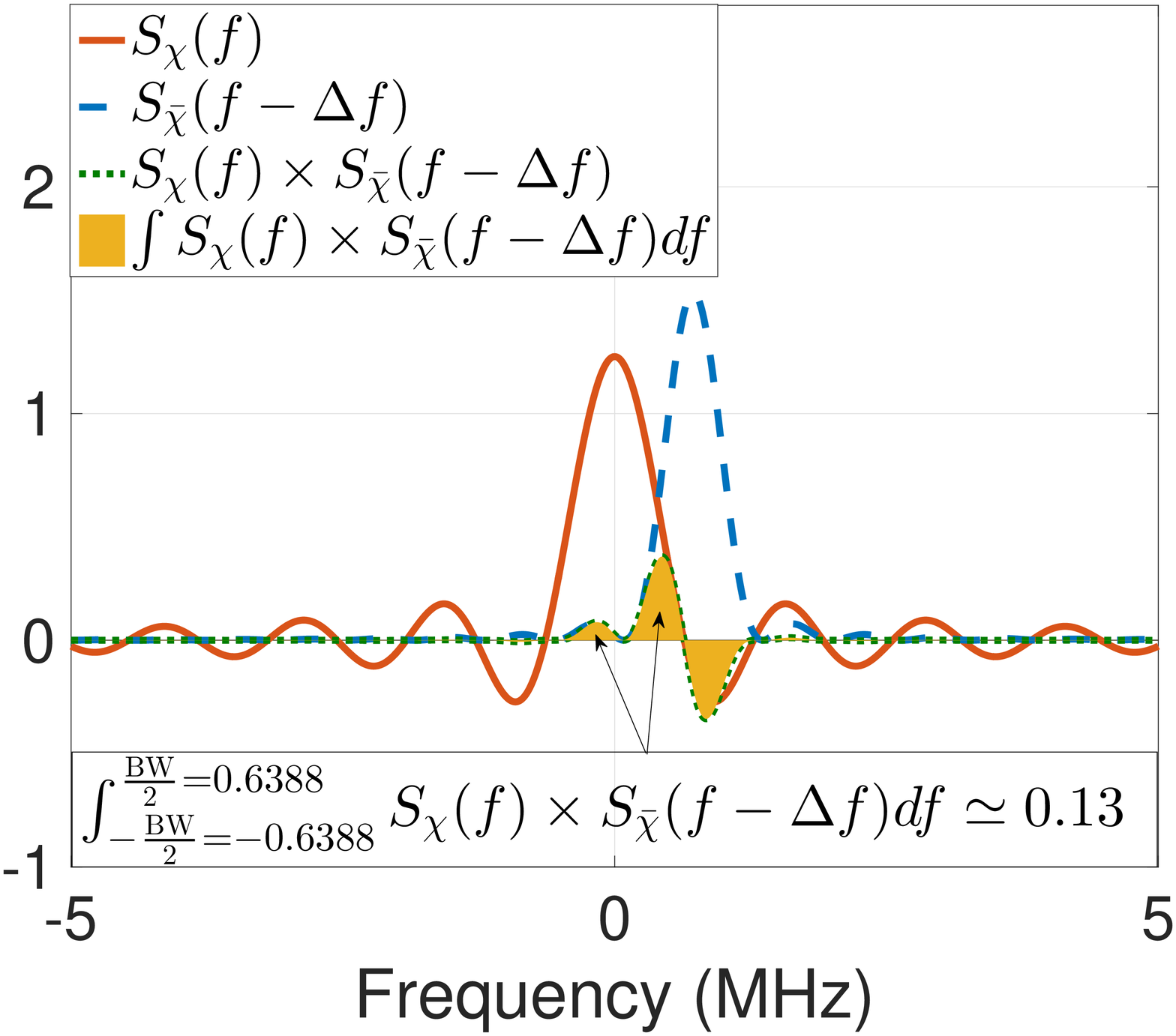}}\caption{\, Frequency domain illustration of the Sinc ($S_{\chi}$) and Sinc$^2$ ($S_{\bar{\chi}}$) pulse shapes at $\alpha_{\rm sp}=0.2776$.}
\label{fig:MinAlpha2}
    \end{subfigure}
    \caption{The effect of pulse-shaping and matched-filtering on the effective interference factors.}
\end{figure*}

\normalsize
To visualize the combined effect of pulse-shaping, duplex parameter $\alpha$, and filtering on the effective cross-mode interference power factor, we plot Fig. \ref{fig:Effective} for different types of pulse shapes. Looking into \eqref{equ:Idd1} and \eqref{equ:Idd2}, along with Fig. \ref{fig:Effective}, several insights on the system operation can be obtained. At HD operation (i.e., $\alpha = 0$) $S_{\chi}$ and $S_{\bar{\chi}}$ exist in non-overlapping null-to-null frequency bands (cf. Fig.~\ref{fig:proposed}), and consequently, $\left|\mathcal{C}_{\chi}(0) \right|^2  \approx 0$. Note that $\left|\mathcal{C}_{\chi}(0) \right|^2 $ is not exactly equal to zero for some pulse shapes at HD operation because of the adjacent channel interference that  exists due to the side ripples of the used pulse shapes. Increasing the duplexing parameter $\alpha$ creates an overlap between the null-to-null frequency bands occupied by $S_{\chi}$ and $S_{\bar{\chi}}$, which increases the effective cross-mode interference power factor $\left|\mathcal{C}_{\chi}(\alpha) \right|^2$. Fig. \ref{fig:Effective} shows that the value of the effective cross-mode interference factor is different for different pulse shapes. A slowly increasing interference factor is desirable as it increases the available BW at a low cost of cross-mode interference. For instance, $\alpha=0.5$ provides 50$\%$ increased BW on the cost of 0.38 cross-mode interference factor when RRC-RRC are used, where this cost can be reduced by $56\%$ if Sinc$^2$-Sinc$^2$ are used. It is worth mentioning that the intra-mode interference is independent of $\alpha$ because all transmitters in the same system use the same pulse shape. 
 
Fig. \ref{fig:Effective} reveals an interesting behavior for the Sinc$^2$-Sinc plus shapes, in which the effective cross-mode interference power factor is nullified at $\alpha =0.2776$. To interpret such behavior, we plot Fig.\ref{fig:MinAlpha}, which shows that an orthogonality between the two pulse shapes occurs for the system using the Sinc$^2$ pulse shape at the specific value of $\alpha = 0.2776$. At this point, the system using the Sinc$^2$ pulse shape gains $0.2776 B$ of the spectrum at no cross-mode interference cost. Note that the orthogonality is not attained for the system using the Sinc pulse shape as shown in Fig. \ref{fig:MinAlpha2}. In this case, the Sinc$^2$ should be assigned to the UL which is more sensitive to the cross-mode interference.

\subsection{LT of the aggregated interference} \label{LTS}

To pursue the analysis and obtain the LT of the interference, we should discriminate between UL and DL cases. In the UL case, the test receiver is a BS and the received SINR at the input of the decoder can be obtained by rewriting \eqref{SINR_chi} as

\small
{\begin{align} \label{SINR_UL}
&  \!\!\!\!\!\!\!\!\!\!\!\!\! {\rm SINR}_{\rm{u}} \left(\alpha \vert \Xi_{\rm{u}}\right) = \frac{ P_{r_o}h_o}{\underset{{k \in \tilde{\Psi}_{\rm u}}}{\sum} P_{{\rm u}_k} h_{{\rm u}_k} r_{{\rm u}_k}^{-\eta} + \underset{j \in \tilde{\Psi}_{\rm d}}{\sum} {{P_{\rm{d}} h_{{\rm d}_j} r_{{\rm d}_j}^{-\eta}} } | \mathcal{\tilde{C}}_{{\rm{u}}}(\alpha)|^2 + {{\beta} P_{\rm{d}} | \mathcal{\tilde{C}}_{{\rm{u}}}(\alpha)|^2} + N_o},
\end{align}}\normalsize
where, $\Xi_{\rm{u}}=\{h_o,P_{r_o}, h_{{\rm u}_k} , r_{{\rm u}_k},P_{{\rm{u}}_k}, h_{{\rm d}_j},  r_{{\rm d}_j}\}$, and $P_{r_o}$ is the received intended signal power. Hence, $P_{r_o}=\rho$ in case the transmitter is CCU and $P_{r_o}=P_{\rm{u}}^{({\rm{M}})} r_o^{-\eta}$ in case of CEU. Due to the employed power control along with the random network topology, the transmit powers of the interfering UEs are random. Let $r_{\rm u}$ be the distance between a CCU and its serving BS, then this UE would transmit with the power $\rho r_{\rm{u}}^{\eta}$. On the other hand, CEUs always transmit at their peak power $P^{(\rm{M})}_{\rm{u}}$. Following \cite{Load2014AlAmmouri}, $P_{{\rm{u}}}$ is a mixed random variable with the probability density function (PDF) given below,
\small
 \begin{align}\label{PowerDistribution}
f_{P_{\rm{u}}}(x)=\left\{
	\begin{array}{ll}
		\frac{2 \pi  \lambda}{\eta \rho^{\frac{2}{\eta}} }x^{\frac{2}{\eta}-1}e^{-\pi  \lambda \left( \frac{x}{\rho}\right)^{\frac{2}{\eta}}}  & \mbox{ } x < P^{({\rm{M}})}_{\rm{u}}.\\
		e^{- \pi  \lambda \left( \frac{P^{({\rm{M}})}_{\rm{u}}}{\rho}\right)^{\frac{2}{\eta}}} & \mbox{ } x= P^{({\rm{M}})}_{\rm{u}}.
	\end{array}
\right.
\end{align}

\normalsize
Equation \eqref{SINR_UL} shows two types of inter-cell interference. The first is the intra-mode inter-cell interference from other UL UEs, which is denoted as $\mathcal{I}_{\rm u \rightarrow u} ={\sum}_{{k \in \tilde{\Psi}_{\rm u}}} P_{{\rm{u}}_i} h_{{\rm u}_{k}} r_{{\rm u}_{k}}^{-\eta}$. The second is the cross-mode inter-cell interference from other DL BSs, which is denoted as $\mathcal{I}_{\rm d \rightarrow u} = {\sum}_{j \in \tilde{\Psi}_{\rm d}} {{P_{\rm{d}} h_{{\rm d}_{j}} r_{{\rm d}_{j}}^{-\eta}} } $. Note that the interference distribution seen by the test UE depends on its type (i.e., CCU or CEU). Therefore, we discriminate between the LT for the interference according to the user type by the superscripts $({\rm CCU})$ and $({\rm CEU})$. Overall, to characterize the UL operation, we need to derive the LT for four different types of interferences, namely, the intra and cross-mode inter-cell interference for CCUs $\mathcal{I}^{({\rm CCU})}_{\rm u \rightarrow u}$, $\mathcal{I}^{({\rm CCU})}_{\rm d \rightarrow u}$, and the intra and cross-mode inter-cell interference for CEUs  $\mathcal{I}^{({\rm CEU})}_{\rm u \rightarrow u}$ and $\mathcal{I}^{({\rm CEU})}_{\rm d \rightarrow u}$.

Due to the universal reuse assumption, the set of interfering BSs is $\tilde{\Psi}_{\rm d} =  {\Psi}_{\rm d} \setminus x_0 $, which is the complete set of BSs excluding the serving BS, hence, $\tilde{\Psi}_{\rm d}$ is a PPP with intensity $ \lambda$. Since each BS can only serve one user at a time on the available channel pair, the intensity of the interfering UEs $\tilde{\Psi}_{\rm u}$ is also  $\lambda$. However, $\tilde{\Psi}_{\rm u}$ is not a PPP due to the employed association technique. That is, only one interfering  UE exists in each Voronoi-cell, which brings correlation among the positions of the interfering UEs and violates the PPP assumption. Furthermore, the employed association makes the set of interfering UEs  $\tilde{\Psi}_{\rm u}$ and the set of interfering BSs  $\tilde{\Psi}_{\rm d}$ dependent. The inter-dependence between the interfering UEs and the cross-dependence between the UEs and BSs impede the model tractability. Hence, to maintain the tractability, we ignore the aforementioned dependencies. The incorporated assumptions to maintain the model tractability are formally stated below.

\begin{assumption}
 The set of interfering UEs $\tilde{\Psi}_{\rm u}$ is a PPP with intensity  $ \lambda$.
 \end{assumption}

\begin{assumption}
The point process $\tilde{\Psi}_{\rm d}$ for the interfering BSs and the point process $\tilde{\Psi}_{\rm u}$ for the interfering UEs are independent.
\end{assumption}

\begin{remark}
 \textbf{Assumption 1} and  \textbf{Assumption 2} are mandatory for the model tractability.  \textbf{Assumption 1} has been used and validated in \cite{Analytical2013Novlan, On2014ElSawy, Hybrid2015Lee, On2014Alves, Load2014AlAmmouri}.  It is important to highlight that both assumptions ignore the mutual correlations between the interfering sources, however, the correlation between the interfering sources and the test receiver is captured through the proper calculation for the interference exclusion region enforced by association and/or  UL power control. The accuracy of the developed model under  \textbf{Assumption 1} and  \textbf{Assumption 2} is validated via independent Monte Carlo simulation in Section~\ref{Results}.
\end{remark}

Exploiting  \textbf{Assumption 1} and  \textbf{Assumption 2}, the LT for the cross-mode and intra-mode interference for CCEs and CEUs are characterized via the following lemma

\begin{lemma}\label{Lem_UL_LT}
The LT of the random variables $\mathcal{I}^{({\rm CCU})}_{\rm d \rightarrow u}$, $\mathcal{I}^{({\rm CCU})}_{\rm u \rightarrow u}$, $\mathcal{I}^{({\rm CEU})}_{\rm d \rightarrow u}$ and $\mathcal{I}^{({\rm CEU})}_{\rm  u \rightarrow u}$ denoted by $\mathcal{L}_{\mathcal{I}^{({\rm CCU})}_{\rm d \rightarrow u}}(s)$, $\mathcal{L}_{\mathcal{I}^{({\rm CCU})}_{\rm  u \rightarrow u}}(s)$, $\mathcal{L}_{\mathcal{I}^{({\rm CEU})}_{\rm d \rightarrow u}}(s)$ and $\mathcal{L}_{\mathcal{I}^{({\rm CEU})}_{\rm  u \rightarrow u}}(s|r_{o})$ are given by the following equations for a general path-loss exponent $\eta$.

\small
\begin{align} \label{LTIuuCCU}
\mathcal{L}_{\mathcal{I}^{({\rm CCU})}_{\rm  u \rightarrow  u}}(s) =\exp \left( - \frac{2 \pi   \lambda}{\eta-2} s \rho^{\frac{-2}{\eta}+1} \mathbb{E}\left[ P_{\rm{u}}^{\frac{2}{\eta}}\right] {}_2 \text{F}_{1}\left(1,1-\frac{2}{\eta},2-\frac{2}{\eta},-s \rho \right) \right),
\end{align}
\begin{align}\label{LTIduCCU}
\mathcal{L}_{\mathcal{I}^{({\rm CCU})}_{\rm d \rightarrow u}}(s)=\mathcal{L}_{\mathcal{I}^{({\rm CEU})}_{\rm d \rightarrow u}}(s)=  \exp\left( - \frac{2}{\eta}\pi^{2}\lambda \left( s P_{\rm{d}} \right)^{\frac{2}{\eta}} \csc \left( \frac{2 \pi}{\eta} \right) \right),
\end{align}
\begin{align}\label{LTIuuCEU}
\mathcal{L}_{\mathcal{I}^{({\rm CEU})}_{\rm  u \rightarrow u}}(s|r_o)=\exp \left(\mathbb{E}_{P_{\rm{u}}}\left[ \frac{-2 \pi  \lambda s P_{\rm{u}} r_o^{2-\eta}}{\eta-2}  {}_2 \text{F}_{1}\left(1,1-\frac{2}{\eta},2-\frac{2}{\eta},-s r_o^{-\eta} P_{\rm{u}}\right)\right] \right),
\end{align}\normalsize
where ${}_2 \text{F}_1 (.)$  is the Hypergeometric function {\rm \cite{Handbook1964Abramowitz}}. The $\frac{2}{\eta}$ fractional moment of the transmit power can be obtained from \eqref{PowerDistribution} as  

\small
\begin{equation}\label{EP}
\mathbb{E}\left[P_{\rm{u}}^\frac{2}{\eta} \right]= \frac{\rho^{\frac{2}{\eta}} \gamma\left(2,\pi \lambda\left(\frac{P^{({\rm{M}})}_{\rm{u}}}{\rho}\right)^{\frac{2}{\eta}}\right)       }{\pi  \lambda} + \left(P^{({\rm{M}})}_{\rm{u}}\right)^{\frac{2}{\eta}} e^{-\pi  \lambda\left(\frac{P^{({\rm{M}})}_{\rm{u}}}{\rho}\right)^{\frac{2}{\eta}} },
\end{equation}
\normalsize
where $\gamma$ is the lower incomplete gamma function {\rm \cite{Handbook1964Abramowitz}}.

Proof: refer to \textbf{Appendix A}.
\end{lemma}

Note that in \eqref{LTIuuCEU} we have to condition on $r_o$ because $r_o$ exists elsewhere in the ${\rm SINR}$ expression (see \eqref{SINR_UL} for $P_{r_o} = P_u^{({\rm M})} r_o^{-\eta}$). On the other hand, the LTs in  \eqref{LTIuuCCU} and \eqref{LTIduCCU} are independent of the service distance $r_o$. Hence, there is no conditioning in \eqref{LTIuuCCU} and \eqref{LTIduCCU}.

In the DL case, the test receiver is a UE and the received SINR at the input of the detector can be obtained by rewriting \eqref{SINR_chi} as
\small
{\begin{align}  \label{SINR_DL}
& {\rm SINR_d} \left(\alpha \vert \Xi_{\rm{d}}\right)  = \frac{P_{\rm{d}} h_o r_o^{-\eta}}{\underset{{k \in \tilde{\Psi}_{\rm d}}}{\sum} {P_{\rm{d}} h_{{\rm d}_k} r_{{{\rm d}_k}}^{-\eta}} + \underset{j \in \tilde{\Psi}_{\rm u}}{\sum} {{P_{{\rm{u}}_j} h_{{\rm u}_j} r_{{\rm u}_j}^{-\eta}} |\mathcal{\tilde{C}}_{\rm{d}}(\alpha)|^2}+ {{\beta} P_{{\rm{u}}_o}  |\mathcal{\tilde{C}}_{\rm{d}}(\alpha)|^2} + N_o},
\end{align}}\normalsize
where $ \Xi_{\rm d} = \{h_o, r_o, h_{{\rm d}_k}, r_{{\rm d}_k}, P_{{\rm{u}}_j}, P_{{\rm{u}}_{o}},  h_{{\rm u}_j}, r_{{\rm u}_j} \}$,  $P_{{\rm{u}}_o}=\rho r_o^{\eta}$ in case the test UE is CCU, and $P_{{\rm{u}}_o}=P_{\rm{u}}^{({\rm{M}})} $ in case the UE is a CEU.

For the DL case and by using similar notations to the UL case and discriminating between the CCU and the CEU performances, we end up with four different types of interferences, namely, the intra and cross-mode inter-cell interference for CCUs $\mathcal{I}^{({\rm CCU})}_{\rm d \rightarrow d}$, $\mathcal{I}^{({\rm CCU})}_{\rm  u \rightarrow d}$, and the intra and cross-mode inter-cell interference for CEUs  $\mathcal{I}^{({\rm CEU})}_{\rm  d \rightarrow d}$ and $\mathcal{I}^{({\rm CEU})}_{\rm  u \rightarrow d}$. Exploiting  \textbf{Assumption 1} and  \textbf{Assumption 2},  the LT for the cross-mode and intra-mode interference for CCEs and CEUs for the DL case are characterized via the following lemma.


\begin{lemma}\label{Lem_DL_LT}
The LT of the random variables $\mathcal{I}^{({\rm CCU})}_{\rm  d \rightarrow d}$, $\mathcal{I}^{({\rm CCU})}_{\rm  u \rightarrow d}$, $\mathcal{I}^{({\rm CEU})}_{\rm  d \rightarrow d}$ and $\mathcal{I}^{({\rm CEU})}_{\rm  u \rightarrow d}$ denoted by $\mathcal{L}_{\mathcal{I}^{({\rm CCU})}_{\rm  d \rightarrow d}}(s|r_{0})$, $\mathcal{L}_{\mathcal{I}^{({\rm CCU})}_{\rm  u \rightarrow d}}(s)$, $\mathcal{L}_{\mathcal{I}^{({\rm CEU})}_{\rm  d \rightarrow d}}(s|r_{0})$ and $\mathcal{L}_{\mathcal{I}^{({\rm CEU})}_{\rm  u \rightarrow d}}(s)$ are given by the following equations for a general path-loss exponent $\eta$ while conditioning on $r_o$.
\small
\begin{align} \label{LTIddCCU}
\!\!\!\!\!\!\!\!\!\!\!  \mathcal{L}_{\mathcal{I}^{({\rm CCU})}_{\rm  d \rightarrow  d}}(s|r_o)=\mathcal{L}_{\mathcal{I}^{({\rm CEU})}_{\rm d \rightarrow d}}(s|r_o) =\exp\left(\frac{-2 \pi  \lambda r_o^{2-\eta} s P_{\rm{d}}}{\eta-2} {}_2 \text{F}_1 \left(1,1-\frac{2}{\eta},2-\frac{2}{\eta},-s P_{\rm{d}} r_o^{-\eta} \right)\right),
\end{align}
\begin{align}\label{LTIudCCU}
\!\!\!\!\!\!\!\!\!\!\! \mathcal{L}_{\mathcal{I}^{({\rm CCU})}_{\rm  u \rightarrow d}}(s)=  \exp\Bigg( &- \frac{2\pi \lambda s  \mathbb{E}\left[{P_{\rm{u}}}^{\frac{2}{\eta}} \right] \rho^{1-\frac{2}{\eta}}  }{\eta-2}  {}_2 \text{F}_1 \left(1,1-\frac{2}{\eta},2-\frac{2}{\eta},-s \rho\right) \Bigg),
\end{align}
\begin{align}\label{LTIudCEU}
\!\!\!\!\!\!\!\!\!\!\!  \mathcal{L}_{\mathcal{I}^{({\rm CEU})}_{\rm  u \rightarrow d}}(s)& =\exp\left( -  \frac{ 2 }{\eta}\pi^{2}  \lambda \csc \left(\frac{2 \pi}{\eta} \right) s^{\frac{2}{\eta}} \mathbb{E}\left[ P_{\rm{u}}^{\frac{2}{\eta}}\right] \right),
\end{align}\normalsize
where $\mathbb{E}\left[ {P_{\rm{u}}}^{\frac{2}{\eta}}\right]$ is given in equation\eqref{EP}.

Proof: refer to  \textbf{Appendix B}.
\end{lemma}

A particular case of interest is at $\eta = 4$, which is a practical value for outdoor communications for cellular networks. In this case, equations in  \textbf{Lemma~\ref{Lem_UL_LT}} and  \textbf{Lemma~\ref{Lem_DL_LT}} reduce to the following forms.

\small
\vspace{-0.5cm}
\begin{align} \label{LTIuuCCU4}
\mathcal{L}_{\mathcal{I}^{({\rm CCU})}_{\rm  u \rightarrow  u}}(s) =\exp\left( - \pi  \lambda \sqrt{s} \mathbb{E}\left[\sqrt{P_{\rm{u}}}  \right]   \arctan\left(\sqrt{s \rho}\right) \right),
\end{align}
\begin{align}\label{LTIduCCU4}
\mathcal{L}_{\mathcal{I}^{({\rm CCU})}_{\rm  d \rightarrow u}}(s)= \mathcal{L}_{\mathcal{I}^{({\rm CEU})}_{\rm d \rightarrow u}}(s)= \exp\left( - \frac{\pi^{2}}{2}  \lambda \sqrt{ s P_{\rm{d}}  } \right),
\end{align}
\begin{align}\label{LTIuuCEU4}
\mathcal{L}_{\mathcal{I}^{({\rm CEU})}_{\rm  u \rightarrow u}}(s|r_o)=\exp \left(- \pi  \lambda \mathbb{E}_{P_{\rm{u}}}\left[ \sqrt{s P_{\rm{u}} } \arctan \left(\sqrt{\frac{s  P_{\rm{u}}}{r_o^4}} \right)  \right] \right),
\end{align}
\begin{align} \label{LTIddCCU4}
\mathcal{L}_{\mathcal{I}^{({\rm CCU})}_{\rm  d \rightarrow  d}}(s|r_o)=\mathcal{L}_{\mathcal{I}^{({\rm CEU})}_{\rm d \rightarrow d}}(s|r_o) =\exp \left( -\pi \lambda \sqrt{s P_{\rm{d}}} \arctan \left( \sqrt{s P_{\rm{d}}} r_o^{-2} \right) \right),
\end{align}
\begin{align}\label{LTIudCCU4}
\mathcal{L}_{\mathcal{I}^{({\rm CCU})}_{\rm  u \rightarrow d}}(s)=  \exp\left( - \pi  \lambda \sqrt{s} \mathbb{E}\left[ \sqrt{P_{\rm{u}}}\right]  \arctan \left(\sqrt{ s\rho} \right) \right),
\end{align}
\begin{align}\label{LTIudCEU4}
\mathcal{L}_{\mathcal{I}^{({\rm CEU})}_{\rm  u \rightarrow d}}(s)& =\exp\left( - \frac{\pi^2 \lambda}{2}    \sqrt{s } \mathbb{E}\left[ \sqrt{P_{\rm{u}}}\right]\right).
\end{align}
\normalsize

Note that the LT in \eqref{LTIuuCEU4} is more computationally complex than the other LTs as it involves an extra averaging step over $P_{\rm{u}}$ inside the exponent. The averaging in \eqref{LTIuuCEU4}  is done w.r.t. the PDF in \eqref{PowerDistribution}. Exploiting Jensen's inequality w.r.t. the random variable $\sqrt{P_{\rm{u}}}$, we can obtain a tight lower-bound simplified bound for \eqref{LTIuuCCU4}, which is given in the following proposition.

\begin{proposition}
The LT of the random variable $\mathcal{I}^{({\rm CEU})}_{\rm  u \rightarrow u}$ which is denoted by $\mathcal{L}_{\mathcal{I}^{({\rm CEU})}_{\rm  u \rightarrow u}}(s|r_{0})$ and given in equation \eqref{LTIuuCEU4} can be simplified by Jensen's inequality which results in the following form.
\small
\begin{align}\label{LTIuuCEU4app}
\mathcal{L}_{\mathcal{I}^{({\rm CEU})}_{\rm  u \rightarrow u}}(s| r_o) < \exp \left(- \pi \lambda \sqrt{s} \mathbb{E}\left[\sqrt{ P_{\rm{u}} }\right]  \arctan \left(\mathbb{E}\left[\sqrt{ P_{\rm{u}} }\right]\sqrt{\frac{s  }{r_o^{4}}} \right)  \right).
\end{align}
\normalsize

Proof: refer to  \textbf{Appendix C}.
\end{proposition}

The bound obtained in \eqref{LTIuuCEU4app} is verified in Section~\ref{Results}. Next, the LTs obtained in this section are used to characterize the BEP, the outage probability  and the transmission rate.

\subsection{\textbf{BEP} Analysis:} \label{Perform1}


Using the SINR expression in \eqref{SINR_DL} and the law of total probability, the DL average BEP for maximum likelihood (ML) decoding with coherent modulation can be obtained by following \cite{The2015Afify} as

 \small
\begin{align}\label{BERDL}
\textbf{BEP}_{\mathcal{DL}}(\alpha)&= \mathbb{E}\left[\omega_1 \text{erfc}\sqrt{\omega_2 \frac{P_{\rm{d}} h_o r_{c}^{-\eta}}{\mathcal{I}^{({\rm CCU})}_{\rm d \rightarrow d} + \mathcal{I}^{({\rm CCU})}_{\rm u \rightarrow d} |\tilde{\mathcal{C}}_{\rm{d}}(\alpha)|^2+ {{\beta} \rho  |\mathcal{\tilde{C}}_{\rm{d}}(\alpha)|^2} + N_o}}\right] {\mathbb{P}\{ \rm{CCU}\}} \notag \\
& \quad \quad \quad \quad +  \mathbb{E}\left[\omega_1 \text{erfc}\sqrt{\omega_2 \frac{P_{\rm{d}} h_o r_{\rm e}^{-\eta}}{\mathcal{I}^{({\rm CEU})}_{\rm d \rightarrow d} + \mathcal{I}^{({\rm CEU})}_{\rm u \rightarrow d} |\tilde{\mathcal{C}}_{\rm{d}}(\alpha)|^2+ {{\beta} P_{\rm u}^{\rm (M)}  |\mathcal{\tilde{C}}_{\rm{d}}(\alpha)|^2} + N_o}}\right] {\mathbb{P}\{{\rm CEU}\}},
\end{align} \normalsize
where, $r_{\rm c}$ and $r_{\rm e}$ are the service distances for CCUs and CEUs respectively, and ${\mathbb{P}\{{\rm CCU}\}}$ and ${\mathbb{P}\{{\rm CEU}\}}$ are the probabilities of being CCU and CEU respectively. Following \cite{Load2014AlAmmouri}, the PDFs of $r_{\rm c}$ and $r_{\rm e}$ and the probabilities of being CCU and CEU are given by
\small
\begin{align}\label{DistanceDistrbutionCCU}
f_{r_{\rm c}}(r)= \frac{2 \pi  \lambda r \exp \left(- \pi  \lambda r^2 \right)}{1- \exp \left(- \pi \lambda {\rm{R}}_{\rm{M}}^2 \right) }\mathbbm{1}_{\left\{0 \le r\le {\rm{R}}_{\rm{M}} \right\}}(r),
\end{align}
\begin{align}\label{DistanceDistrbutionCEU}
f_{r_{\rm e}}(r)= 2 \pi  \lambda r \exp \left(- \pi  \lambda r^2+\pi  \lambda {\rm{R}}_{\rm{M}}^2  \right) \mathbbm{1}_{\left\{{\rm{R}}_{\rm{M}} < r \le \infty \right\}}(r),
\end{align}
\begin{align}\label{PrCCU}
\mathbb{P}\{{\rm CCU}\}&= 1-\exp \left( - \pi \lambda \rm{R^2_M}\right).
\end{align}
\begin{align}\label{PrCEU}
\mathbb{P}\{{\rm CEU}\}&= \exp \left( - \pi \lambda \rm{R^2_M}\right).
\end{align}
\normalsize
\noindent where ${\rm{R}}_{\rm{M}}= \left( \frac{P^{({\rm{M}})}_{\rm{u}}}{\rho} \right)^{\frac{1}{\eta}}$ is the boundary distance between CCUs and CEUs.  

Similarly, using the SINR expression in \eqref{SINR_UL}, the UL average BEP is given by
\small
 \begin{align}\label{BERUL1}
\textbf{BEP}_{\mathcal{UL}}(\alpha)&= \mathbb{E}\left[\omega_1 \text{erfc}\sqrt{\omega_2 \frac{\rho h_o }{\mathcal{I}^{({\rm CCU})}_{\rm u \rightarrow u} + \mathcal{I}^{({\rm CCU})}_{\rm d \rightarrow u} |\tilde{\mathcal{C}}_{\rm{u}}(\alpha)|^2+ {{\beta} P_{\rm{d}}  |\mathcal{\tilde{C}}_{\rm{u}}(\alpha)|^2} + N_o} }\right] {\mathbb{P}\{ \rm{CCU}\}} \notag \\
& \quad \quad   +  \mathbb{E}\left[\omega_1 \text{erfc}\sqrt{\omega_2 \frac{P_{\rm{u}}^{(\rm{M})} h_o r_{\rm e}^{-\eta}}{\mathcal{I}^{({\rm CEU})}_{\rm u \rightarrow u} + \mathcal{I}^{({\rm CEU})}_{\rm d \rightarrow u} |\tilde{\mathcal{C}}_{\rm{u}}(\alpha)|^2+ {{\beta} P_{\rm{d}}}  |\mathcal{\tilde{C}}_{\rm{u}}(\alpha)|^2 + N_o}}\right] {\mathbb{P}\{{\rm CEU}\}}.
\end{align} \normalsize
 
Following \cite{A2009Shobowale}, the expectations in the form of  \eqref{BERDL} and \eqref{BERUL1} can be evaluated in terms of the LT of the interference. Hence, using \textbf{Lemma~\ref{Lem_UL_LT}} and \textbf{Lemma~\ref{Lem_DL_LT}}, the average BEP for the depicted system model is given in the following theorem.

\begin{theorem}
\label{theorem1}
In a single-tier Poisson cellular network with channel inversion power control of threshold $\rho$, $2\alpha B$ overlap between the UL and DL frequency bands, and exponentially distributed channel gains with unity means, the BEP in the UL and DL directions for a generic user and a generic BS can be found by the following equations.

\scriptsize
\begin{align}\label{BEdGenCCU}
  \textbf{BEP}_{\mathcal{DL}}(\alpha) =\omega_1^{({\rm{d}})} -& \mathbb{P}\left\{CCU\right\} \int\limits_{0}^{\rm{R_M}}  \int\limits_{0}^{\infty}\frac{ \omega_1^{({\rm{d}})}f_{r_{\rm c}}(r)\mathcal{L}_{\mathcal{I}^{({\rm CCU})}_{\rm d \rightarrow  d}}\left(\frac{z r^{\eta}}{ P_{\rm{d}} \omega_2^{({\rm{d}})}} \right) \mathcal{L}_{\mathcal{I}^{({\rm CCU})}_{\rm u \rightarrow  d}}\left(\frac{r^{\eta}z|\tilde{\mathcal{C}}_{\rm{d}}(\alpha)|^2}{P_{\rm{d}} \omega_2^{({\rm{d}})}}\right) } { \sqrt{\pi z} } \notag \\ 
&\times \exp \Bigg(-z\left(1+ \frac{{\beta} \rho  |\mathcal{\tilde{C}}_{\rm{d}}(\alpha)|^2 r^{2 \eta}}{\omega_2^{({\rm{d}})} P_{\rm{d}}  } + \frac{N_o r^{\eta}}{{ \omega_2^{({\rm{d}})}  P_{\rm{d}} }}\right) \Bigg)dz \ dr \notag \\
&  - \mathbb{P}\left\{CEU\right\} \int\limits_{\rm{R_M}}^{\infty} \int\limits_{0}^{\infty} \frac{\omega_1^{({\rm{d}})} f_{r_{\rm e}}(r)\mathcal{L}_{\mathcal{I}^{({\rm CEU})}_{\rm d \rightarrow  d}}\left(\frac{z r^{\eta}}{ P_{\rm{d}} \omega_2^{({\rm{d}})}}\right) \mathcal{L}_{\mathcal{I}^{({\rm CEU})}_{\rm u \rightarrow  d}}\left(\frac{z r^{\eta}|\tilde{\mathcal{C}}_{\rm{d}}(\alpha)|^2}{P_{\rm{d}} \omega_2^{({\rm{d}})}}\right) } {\sqrt{\pi z} } \notag \\
& \times \exp \Bigg(-z\left(1+ \frac{{\beta} P_{\rm{u}}^{({\rm{M}})}  |\mathcal{\tilde{C}}_{\rm{d}}(\alpha)|^2  r^{\eta}}{\omega_2^{({\rm{d}})} P_{\rm{d}} } + \frac{N_o r^{\eta}}{{\omega_2^{({\rm{d}})}  P_{\rm{d}} }}\right)  \Bigg)  dz \ dr,
\end{align}

\begin{align}\label{BEuGenCCU}
  \!\!\!\!\!\!\!\!\!\!\!\!   \textbf{BEP}_{\mathcal{UL}}(\alpha)=\omega_1^{({\rm{u}})} -  & \mathbb{P}\left\{CCU\right\} \int\limits_{0}^{\infty} \frac{\omega_1^{({\rm{u}})}\mathcal{L}_{\mathcal{I}^{({\rm CCU})}_{\rm u \rightarrow  u}} \left(\frac{z}{\rho \omega_2^{({\rm{u}})}}\right) \mathcal{L}_{\mathcal{I}^{({\rm CCU})}_{\rm d \rightarrow  u}} \left(\frac{z |\tilde{\mathcal{C}}_{\rm{u}}(\alpha)|^2}{\rho \omega_2^{({\rm{u}})}} \right) }{\sqrt{\pi z} } \exp \Bigg(-z\left(1+\frac{{\beta} P_{\rm{d}}  |\mathcal{\tilde{C}}_{\rm{u}}(\alpha)|^2}{{\omega_2^{({\rm{u}})} \rho}} + \frac{N_o}{{ \omega_2^{({\rm{u}})} \rho}}\right) \Bigg) dz \notag \\
    &-  \mathbb{P}\left\{CEU\right\} \int\limits_{R_M}^{\infty} \int\limits_{0}^{\infty}  \frac{\omega_1^{({\rm{u}})} f_{r_{\rm e}}(r)\mathcal{L}_{\mathcal{I}^{({\rm CEU})}_{\rm u \rightarrow  u}} \left(\frac{z r^{\eta}}{P_{\rm{u}}^{(\rm{M})} \omega_2^{({\rm{u}})}}\right) \mathcal{L}_{\mathcal{I}^{({\rm CEU})}_{\rm d \rightarrow  u}}\left(\frac{r^{\eta} z|\tilde{\mathcal{C}}_{\rm{u}}(\alpha)|^2}{P_{\rm{u}}^{(\rm{M})} \omega_2^{({\rm{u}})}} \right) }{\sqrt{\pi z} } \notag \\  
    &\times \exp \Bigg(-z\left(1+\frac{{\beta} P_{\rm{d}}  |\mathcal{\tilde{C}}_{\rm{d}}(\alpha)|^2 r^{\eta}}{{\omega_2^{({\rm{u}})}  P_{\rm{u}}^{({\rm{M}})}}} + \frac{N_o r^{\eta}}{{ \omega_2^{({\rm{u}} )} P_{\rm{u}}^{({\rm{M}})}}}\right) \Bigg)  dz \ dr,
\end{align}\normalsize
where, $f_{r_{\rm c}}(r)$, $f_{r_{\rm e}}(r)$, $\mathbb{P}\left\{\rm CCU \right\}$, and   $\mathbb{P}\left\{\rm CEU \right\}$, are given in \eqref{DistanceDistrbutionCCU}, \eqref{DistanceDistrbutionCEU}, \eqref{PrCCU}, and \eqref{PrCEU}, respectively. The LT of the cross-mode and intra-mode interference in the UL and DL are given in \textbf{Lemma~\ref{Lem_UL_LT}} and \textbf{Lemma~\ref{Lem_DL_LT}}.

Proof: see \textbf{Appendix D}.
\end{theorem}

\subsection{Outage Probability:} \label{Perform2}

A simpler, but more abstract, technique to asses transmission reliability is to look at the outage probability. The outage probability is defined as the probability that the SINR falls below a certain threshold $\theta$, which can give closed from expressions in some special cases. The outage probabilities for the DL and the UL are expressed as
\small
\begin{align}\label{OutageDL}
\!\!\!\!\!\!\!\!\! \mathcal{O}_{\mathcal{DL}}(\alpha, \theta)&=   \mathbb{P}\left\{ \frac{P_{\rm{d}} h_o r_{c}^{-\eta}}{\mathcal{I}^{({\rm CCU})}_{\rm d \rightarrow d} + \mathcal{I}^{({\rm CCU})}_{\rm u \rightarrow d} |\tilde{\mathcal{C}}_{\rm{d}}(\alpha)|^2+ {{\beta} \rho  |\mathcal{\tilde{C}}_{\rm{d}}(\alpha)|^2} + N_o} < \theta \right\} {\mathbb{P}\{ \rm{CCU}\}} \notag \\
& \quad \quad  +  \mathbb{P}\left\{ \frac{P_{\rm{d}} h_o r_{\rm e}^{-\eta}}{\mathcal{I}^{({\rm CEU})}_{\rm d \rightarrow d} + \mathcal{I}^{({\rm CEU})}_{\rm u \rightarrow d} |\tilde{\mathcal{C}}_{\rm{d}}(\alpha)|^2+ {{\beta} P_{\rm u}^{\rm (M)}  |\mathcal{\tilde{C}}_{\rm{d}}(\alpha)|^2} + N_o} <  \theta \right\} {\mathbb{P}\{{\rm CEU}\}},
\end{align} 

 \begin{align}\label{BERUL}
\!\!\!\!\!\!\!\!\! \mathcal{O}_{\mathcal{UL}}(\alpha, \theta)&= \mathbb{P}\left\{ \frac{\rho h_o }{\mathcal{I}^{({\rm CCU})}_{\rm u \rightarrow u} + \mathcal{I}^{({\rm CCU})}_{\rm d \rightarrow u} |\tilde{\mathcal{C}}_{\rm{u}}(\alpha)|^2+ {{\beta} P_{\rm{d}}  |\mathcal{\tilde{C}}_{\rm{u}}(\alpha)|^2} + N_o}  < \theta\right\} {\mathbb{P}\{ \rm{CCU}\}} \notag \\
& \quad \quad   +  \mathbb{P}\left\{  \frac{P_{\rm u}^{\rm (M)} h_o r_{\rm e}^{-\eta}}{\mathcal{I}^{({\rm CEU})}_{\rm u \rightarrow u} + \mathcal{I}^{({\rm CEU})}_{\rm d \rightarrow u} |\tilde{\mathcal{C}}_{\rm{u}}(\alpha)|^2+ {{\beta} P_{\rm{d}}}  |\mathcal{\tilde{C}}_{\rm{u}}(\alpha)|^2 + N_o} < \theta \right\}  {\mathbb{P}\{{\rm CEU}\}},
\end{align} \normalsize
where, ${\mathbb{P}\{{\rm CCU}\}}$ and ${\mathbb{P}\{{\rm CEU}\}}$ are given in \eqref{PrCCU} and \eqref{PrCEU}, respectively. The outage probability for the UL and DL cases are given in the following theorem.

\begin{theorem}
\label{theorem2}
In a single-tier Poisson cellular network with channel inversion power control of threshold $\rho$, $2\alpha B$ overlap between the UL and DL frequency bands, and exponentially distributed channel gains with unity means, the outage probability in the UL and DL directions for a generic user and a generic BS are characterized by the following equations.
\scriptsize

\begin{align}\label{OutdGenCCU}
& \!\!\!\!\!\!\!\mathcal{O}_{\mathcal{DL}}(\alpha,\theta)=1-  {\mathbb{P}\{{\rm CCU}\}} \int\limits_{0}^{R_M}  f_{r_{\rm c}}(r)\mathcal{L}_{\mathcal{I}^{({\rm CCU})}_{\rm d \rightarrow  d}}\left(\frac{\theta r^{\eta}}{P_{\rm{d}}}\right) \mathcal{L}_{\mathcal{I}^{({\rm CCU})}_{\rm u \rightarrow  d}}\left(\frac{\theta r^{\eta}|\tilde{\mathcal{C}}_{\rm{d}}(\alpha)|^2}{P_{\rm{d}}}\right)   \exp \left(- \theta \left(  \frac{{\beta} \rho  |\mathcal{\tilde{C}}_{\rm{d}}(\alpha)|^2 r^{2 \eta}}{ P_{\rm{d}}  } +\frac{N_o r^{\eta}}{{  P_{\rm{d}} }}  \right) \right) \ dr  \notag \\
& \quad \quad \quad- {\mathbb{P}\{{\rm CEU}\}} \int\limits_{\rm{R_M}}^{\infty} f_{r_{\rm e}}(r)\mathcal{L}_{\mathcal{I}^{({\rm CEU})}_{\rm d \rightarrow  d}}\left(\frac{\theta r^{\eta}}{P_{\rm{d}}}\right) \mathcal{L}_{\mathcal{I}^{({\rm CEU})}_{\rm u \rightarrow  d}}\left(\frac{\theta r^{\eta}|\tilde{\mathcal{C}}_{\rm{d}}(\alpha)|^2}{P_{\rm{d}}}\right)  \exp \left(-\theta  \left(   \frac{{\beta} P_{\rm{u}}^{({\rm{M}})}  |\mathcal{\tilde{C}}_{\rm{d}}(\alpha)|^2  r^{\eta}}{ P_{\rm{d}} } + \frac{N_o r^{\eta}}{{  P_{\rm{d}} }} \right)  \right)   \ dr,
\end{align}

\begin{align}\label{OutuGenCCU}
& \!\!\!\!\!\!\!  \mathcal{O}_{\mathcal{UL}}(\alpha, \theta)=1- \mathcal{L}_{\mathcal{I}^{({\rm CCU})}_{\rm u \rightarrow  u}}\left(\frac{\theta }{\rho}\right) \mathcal{L}_{\mathcal{I}^{({\rm CCU})}_{\rm d \rightarrow  u}}\left(\frac{\theta |\tilde{\mathcal{C}}_{\rm{u}}(\alpha)|^2 }{\rho} \right) \exp \Bigg(- \theta \left(\frac{{\beta} P_{\rm{d}}  |\mathcal{\tilde{C}}_{\rm{u}}(\alpha)|^2}{ \rho} + \frac{N_o}{ \rho}\right) \Bigg) \notag \\
& \!\!\!\!\!\!\! - \int\limits_{\rm{R_M}}^{\infty}  f_{r_{\rm e}}(r)\mathcal{L}_{\mathcal{I}^{({\rm CEU})}_{\rm u \rightarrow  u}}\left(\frac{\theta r^{\eta} }{P_{\rm{u}}^{({\rm{M}})}} \right) \mathcal{L}_{\mathcal{I}^{({\rm CEU})}_{\rm d \rightarrow  u}}\left(\frac{\theta r^{\eta} |\tilde{\mathcal{C}}_{\rm{u}}(\alpha)|^2}{P_{\rm{u}}^{({\rm{M}})}} \right) \exp \Bigg(-\theta \left(\frac{{\beta} P_{\rm{d}}  |\mathcal{\tilde{C}}_{\rm{u}}(\alpha)|^2 r^{\eta}}{{  P_{\rm{u}}^{({\rm{M}})}}} + \frac{N_o r^{\eta}}{{  P_{\rm{u}}^{({\rm{M}})}}}\right) \Bigg) \ dr,
\end{align}\normalsize
where,  $f_{r_{\rm c}}(r)$, $f_{r_{\rm e}}(r)$, $\mathbb{P}\left\{\rm CCU \right\}$, and   $\mathbb{P}\left\{\rm CEU \right\}$, are given in \eqref{DistanceDistrbutionCCU}, \eqref{DistanceDistrbutionCEU}, \eqref{PrCCU}, and \eqref{PrCEU}, respectively. The LT of the cross-mode and intra-mode interference in the UL and the DL are given in \textbf{Lemma~\ref{Lem_UL_LT}} and \textbf{Lemma~\ref{Lem_DL_LT}}.
\small

\normalsize
Proof: see \textbf{Appendix E}.
\end{theorem}

To get simple expressions for the outage probability, we consider the special case of $\eta=4$, interference limited, and unbinding UL transmit power. Note that the unbinding UL transmit power can be interpreted as the case where the BSs are sufficiently dense such that all users can invert their channels, which is a common case in urban areas and downtowns \cite{On2014ElSawy}. Also, the interference limited case is a reasonable assumption given that cellular networks are interference limited. In this case, the equations in \textbf{Theorem 2} reduce to

\small
\begin{align}\label{OutdSpecial}
& \!\!\!\!\!\!\!\mathcal{O}_{\mathcal{DL}}(\alpha,\theta)=1- \int\limits_{0}^{\infty} 2 \pi \lambda r \exp \left(-\pi \lambda r^2- \pi \lambda {\rm U}(\theta)r^2 - {\rm U}\left(\frac{\rho \theta r^4 |\mathcal{\tilde{C}}_{\rm{d}}(\alpha)|^2}{P_{\rm d}} \right)-\theta \frac{{\beta} \rho  |\mathcal{\tilde{C}}_{\rm{d}}(\alpha)|^2 r^{8}}{ P_{\rm{d}}  }\right) dr,
\end{align}
\begin{align}\label{OutuSpecial2}
& \!\!\!\!\!\!\!\mathcal{O}_{\mathcal{UL}}(\alpha,\theta)=1- \exp \left( - {\rm U}(\theta)-\frac{\pi^2}{2} \lambda \sqrt{\frac{\theta P_{\rm d}}{\rho}} |\mathcal{\tilde{C}}_{\rm{u}}(\alpha)|-\theta \frac{{\beta} P_{\rm{d}}  |\mathcal{\tilde{C}}_{\rm{u}}(\alpha)|^2}{ \rho} \right),
\end{align}
\normalsize

\noindent where ${\rm U}(x)=\sqrt{x} \arctan(\sqrt{x})$. In the case of perfect SIC\footnote{Assuming perfect SIC captures the explicit effect of the cross-mode interference, which is the main focus of the paper.}, a closed form approximation for equation \eqref{OutdSpecial} can be derived by using the $\arctan(\cdot)$ approximation in \cite[Equation 2]{Efficient2006Rajan} which leads to

\small
\begin{align}\label{OutdSpecialApp}
& \!\!\!\!\!\!\!\mathcal{O}_{\mathcal{DL}}(\alpha,\theta) \approx 1-\frac{\pi ^{3/2} \lambda  \exp \left(\frac{\pi ^2 \lambda ^2 P_{\rm d} ({\rm U} (\theta)+1)^2}{4 |\mathcal{\tilde{C}}_{\rm{d}}(\alpha)|^2 \theta  \rho }\right) \text{erfc}\left(\frac{\pi  \lambda  ({\rm U} (\theta)+1)}{2 \sqrt{\frac{|\mathcal{\tilde{C}}_{\rm{d}}(\alpha)|^2 \theta  \rho }{P_{\rm d}}}}\right)}{2 \sqrt{\frac{|\mathcal{\tilde{C}}_{\rm{d}}(\alpha)|^2 \theta  \rho }{P_{\rm d}}}}.
\end{align}\normalsize

\noindent where ${\rm erfc(\cdot)}$ is the complementary error function \cite[Equation 7.1.2]{Handbook1964Abramowitz}.

\subsection{Transmission Rate} \label{Perform3}

In this section, we obtain the transmission rate in two cases. First, assuming adaptive rate transmission with perfect knowledge of the channel state information (CSI), denoted as ergodic rate \cite{On2014ElSawy,A2011Andrews}. Second, assuming fixed transmission rate with unknown CSI, denoted as effective rate used in \cite{Throughput2014Li}. In the former case, the ergodic rate $\left(\mathcal{R}(\alpha)\right)$ is defined as $\mathbb{E}\left[{{\rm BW}}(\alpha) \log_2 \left(1+{\rm SINR}(\alpha) \right)\right]$ which can be expressed as
 
\small
\begin{align}\label{TotalRate}
\mathcal{R}(\alpha)=\mathbb{E}\left[{{\rm BW}}(\alpha) \log_2 \left(1+{\rm SINR}(\alpha) \right)\right] &\stackrel{(i)}{=}\int\limits_{0}^{\infty} \mathbb{P}\left\{{{\rm BW}}(\alpha) \log_2 \left(1+{\rm SINR}(\alpha) \right)>t \right\} dt, \notag \\
&\stackrel{(ii)}{=} \int\limits_{0}^{\infty} \left[1- \mathcal{O}(\alpha, 2^{\frac{t}{{\rm BW}(\alpha)}}-1) \right] dt,
\end{align}\normalsize
where the SINR is defined in \eqref{SINR_UL} and \eqref{SINR_DL} for the UL and DL schemes, respectively. In \eqref{TotalRate} $(i)$ follows from the fact that the ${\rm SINR}$ is a positive random variable and $(ii)$ from the definition of the outage probability. On the other hand, the effective rate ($\mathcal{E}(\alpha,\theta)$) is defined as ${\rm BW}(\alpha) \  \mathbb{P} \left\{ {\rm SINR}(\alpha)>\theta \right\} \log_2 (1+\theta)$ where the nodes transmit with a fixed rate ${\rm BW(\alpha)} \  \log_2 (1+\theta)$ regardless of the state of the channel. Hence, the rate of the successfully transmitted symbols is given by

\begin{align}\label{EffectiveRate}
\mathcal{E}(\alpha,\theta)= {\rm BW}(\alpha) \log_2(1+\theta) (1-\mathcal{O}(\alpha,\theta)). 
\end{align}\normalsize

We look at the interesting special cases that leads to simple expressions for the ergodic and effective rates. General expression are omitted due to space constraints and can be directly obtained by substituting the outage expressions from Theorem 2 in \eqref{TotalRate} and \eqref{EffectiveRate}, respectively. An interesting special case that gives simplified expressions is presented in the following proposition.

\begin{proposition}
Assuming interference limited network (i.e., ignoring noise), $\eta$=4, and unbinding UL transmit power, the ergodic and effective rates are given by,

\small
\begin{align}
& \! \! \! \! \! \! \! \! \! \! \! \!\mathcal{R}_{UL}(\alpha)= \int\limits_{0}^{\infty} \frac{B_{\rm{u}}+\alpha B}{\ln(2) (g+1) } \exp \left(- U(g) - \frac{\pi^2 \lambda }{2} \sqrt{\frac{g}{\rho}}  \sqrt{P_{\rm{d}} |\mathcal{\tilde{C}}_{\rm{u}}(\alpha)|^2} -g \frac{{\beta} P_{\rm{d}}  |\mathcal{\tilde{C}}_{\rm{u}}(\alpha)|^2}{ \rho}\right)dg,
\end{align}
\begin{align}
& \! \! \! \! \! \!\! \! \! \! \! \! \mathcal{R}_{DL}(\alpha) = \int\limits_{0}^{\infty} \int\limits_{0}^{\infty}  \frac{2 \pi \lambda r \left(B_{\rm{d}}+\alpha B \right)}{\ln(2) (g+1) }  \exp \left(-\pi \lambda r^2- \pi \lambda {\rm U}(g)r^2 - {\rm U}\left(\frac{\rho g r^4 |\mathcal{\tilde{C}}_{\rm{d}}(\alpha)|^2}{P_{\rm d}} \right)-g \frac{{\beta} \rho  |\mathcal{\tilde{C}}_{\rm{d}}(\alpha)|^2 r^{8}}{ P_{\rm{d}}  }\right) dr  dg,
\end{align}
\begin{align} \label{the_only}
& \! \! \! \! \! \!\! \! \! \! \! \! \mathcal{E}_{UL}(\alpha,\theta)= \left( B_{\rm{u}}+\alpha B \right) \log_2 (1+\theta) \exp \left(- U(\theta) - \frac{\pi^2 \lambda }{2} \sqrt{\frac{\theta}{\rho}}  \sqrt{P_{\rm{d}} |\mathcal{\tilde{C}}_{\rm{u}}(\alpha)|^2} -\theta \frac{{\beta} P_{\rm{d}}  |\mathcal{\tilde{C}}_{\rm{u}}(\alpha)|^2}{ \rho}\right),
\end{align}
\begin{align}
& \! \! \! \! \! \!\! \! \! \! \! \! \mathcal{E}_{DL}(\alpha,\theta) =  \left(B_{\rm{d}}+\alpha B \right)\log_2 (1+\theta)\int\limits_{0}^{\infty} 2 \pi \lambda r \exp \left(-\pi \lambda r^2- \pi \lambda {\rm U}(\theta)r^2 - {\rm U}\left(\frac{\rho \theta r^4 |\mathcal{\tilde{C}}_{\rm{d}}(\alpha)|^2}{P_{\rm d}} \right)-\theta \frac{{\beta} \rho  |\mathcal{\tilde{C}}_{\rm{d}}(\alpha)|^2 r^{8}}{ P_{\rm{d}}  }\right) dr,
\end{align}\normalsize
where $U(\cdot)$ is given in equations \eqref{OutdSpecial} and \eqref{OutuSpecial2}. The DL rates can be further simplified by assuming perfect SIC and using the outage probability approximation given by equation \eqref{OutdSpecialApp} which result in

\small
\begin{align} \label{the_onlu_Dl1}
& \! \! \! \! \! \! \mathcal{R}_{DL}(\alpha) \approx \int\limits_{0}^{\infty}  \frac{\pi ^{3/2} \lambda  \left(B_{\rm{d}}+\alpha B \right)}{2\ln(2) (g+1) \sqrt{\frac{|\mathcal{\tilde{C}}_{\rm{d}}(\alpha)|^2 g  \rho }{P_{\rm d}}} }   \exp \left(\frac{\pi ^2 \lambda ^2 P_{\rm d} ({\rm U} (g)+1)^2}{4 |\mathcal{\tilde{C}}_{\rm{d}}(\alpha)|^2 g  \rho }\right) \text{erfc}\left(\frac{\pi  \lambda  ({\rm U} (g)+1)}{2 \sqrt{\frac{|\mathcal{\tilde{C}}_{\rm{d}}(\alpha)|^2 g  \rho }{P_{\rm d}}}}\right)  dg,
\end{align}
\begin{align} \label{the_onlu_Dl2}
& \! \! \! \! \! \! \mathcal{E}_{DL}(\alpha,\theta) \approx\frac{\pi ^{3/2} \lambda  \left(B_{\rm{d}}+\alpha B \right)\log_2 (1+\theta) }{ \sqrt{\frac{|\mathcal{\tilde{C}}_{\rm{d}}(\alpha)|^2 \theta  \rho }{P_{\rm d}}}}  \exp \left(\frac{\pi ^2 \lambda ^2 P_{\rm d} ({\rm U} (\theta)+1)^2}{ |\mathcal{\tilde{C}}_{\rm{d}}(\alpha)|^2 \theta  \rho }\right) \text{erfc}\left(\frac{\pi  \lambda  ({\rm U} (\theta)+1)}{ \sqrt{\frac{|\mathcal{\tilde{C}}_{\rm{d}}(\alpha)|^2 \theta  \rho }{P_{\rm d}}}}\right),
\end{align}\normalsize
\end{proposition}

The approximations in Proposition 2 are verified in Section~\ref{Results}.

\subsection{Discussion}  \label{dicsus}

\begin{figure}[t]
\centerline{\includegraphics[width=  2.2in]{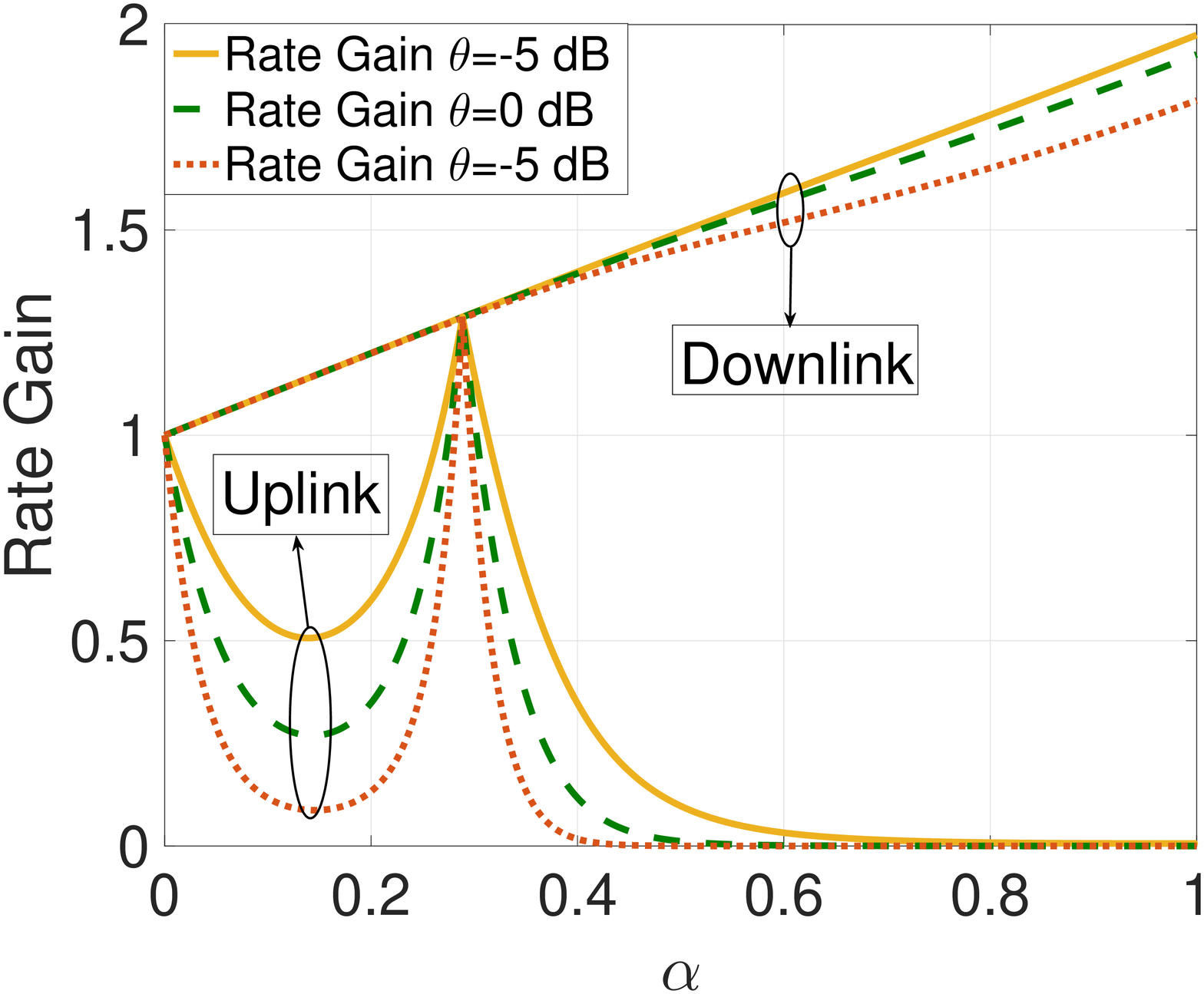}}\caption{\, UL and DL rate gains vs the duplexing parameter $\alpha$ with perfect SIC ($\beta=0$).}
\label{fig:Gains}
\vspace{-0.8cm}
\end{figure}

Before looking at numerical and simulation results, we discuss some insights that can be directly deduced from the developed mathematical model. Particularly, we highlight the vulnerability of the UL to cross-mode interference and look at the $\alpha$-duplex to HD rate ratio to show the required conditions to achieve rate gains form the $\alpha$-duplex scheme.

The vulnerability of the UL to the cross-mode interference is revealed from \eqref{the_only}, in which the rate linearly increases in $\alpha$ but exponentially degrades with the cross-mode interference factor. In the case of perfect SIC, the sensitivity of the UL rate to the  cross-mode interference factor depends on the term $\lambda \sqrt{\frac{P_{\rm d}}{\rho}}$, which is typically large due to the high BSs transmit power and the low target received power level of the employed UL power control.

The UL rate gain is defined as  $\mathcal{G}(\alpha,\theta)_{UL} = \frac{\mathcal{E}_{UL}(\alpha,\theta)}{\mathcal{E}_{UL}(0,\theta)}$, where $ \mathcal{E}_{UL}(\alpha,\theta)$ is defined by \eqref{the_only}. It can be shown that the condition for achieving $\mathcal{G}(\alpha,\theta)_{UL}>1$ assuming perfect SIC is

\small
\begin{align} \label{the_only2}
 \sqrt\frac{\rho}{P_{\rm d}} &>  \frac{\pi^2 \lambda \theta}{(2 \ln (1+\alpha))}   (|\mathcal{\tilde{C}}_{\rm{u}}(\alpha)|-|\mathcal{\tilde{C}}_{\rm{u}}(0)|),
\end{align}\normalsize
which is trivially satisfied at pulse-shapes orthogonality (i.e., $|\mathcal{\tilde{C}}_{\rm{u}}(\alpha)|=0$) and hard to achieve at FD operation (i.e., $\alpha =1$). For the sake of illustration, we consider a numerical example with rectangular (in frequency domain) pulses for which $ (|\mathcal{\tilde{C}}_{\rm{u}}(1)|-|\mathcal{\tilde{C}}_{\rm{u}}(0)| =1$. In this case, looking at FD operation with 0 dB, i.e., $\theta=1$, decoding threshold, \eqref{the_only2} reduces to 

\small
\begin{align}\label{NewRateuSpecial}
 \sqrt{\frac{\rho}{P_{\rm d}}} &> \left(\frac{\pi^2 \lambda}{2\ln (2) }\right) = 7.1194 \lambda
\end{align}\normalsize
which is hard to achieve due to the low value of $\frac{\rho}{P_{\rm d}}$ in addition to the relatively high value of $\lambda$ that satisfies the unbinding UL power control condition, under-which \eqref{NewRateuSpecial} holds. Note that the derived condition is hard to achieve given that perfect SIC is assumed, so for practical imperfect SIC it is even harder to achieve. 

For the DL case, the rate gain, defined as $\mathcal{G}_{DL}(\alpha,\theta)=\mathbb{E}_{r} \left[\frac{\mathcal{E}_{DL}(\alpha,\theta |r)}{\mathcal{E}_{DL}(0,\theta|r)}\right]$ assuming perfect SIC, is given by

\small
\begin{align} \label{the_onlu_Dl3}
& \! \! \! \! \! \! \mathcal{G}_{DL}(\alpha,\theta) \approx \frac{\pi ^{3/2} \lambda  (1+\alpha) }{ \sqrt{\frac{|\mathcal{\tilde{C}}_{\rm{d}}(\alpha)|^2 \theta  \rho }{P_{\rm d}}}  }  \exp \left(\frac{\pi ^2 \lambda ^2 P_{\rm d} }{4  |\mathcal{\tilde{C}}_{\rm{d}}(\alpha)|^2 \theta  \rho }\right) \text{erfc}\left(\frac{\pi  \lambda  }{2 \sqrt{\frac{ |\mathcal{\tilde{C}}_{\rm{d}}(\alpha)|^2 \theta  \rho }{P_{\rm d}}}}\right),
\end{align}\normalsize
which is hard to be analytically proven to be greater than unity. Hence, we plot the rate gains $\mathcal{G}_{UL}(\alpha,\theta)$ and $\mathcal{G}_{DL}(\alpha,\theta)$ in Fig. \ref{fig:Gains}. The figure shows that the DL always benefits from increasing $\alpha$, which shows the immunity of the DL to the UL interference. The figure also emphasizes the sensitivity of the UL to the DL interference and shows that the condition in \eqref{the_only2} is not satisfied away form the orthogonality point (i.e., $\alpha = \alpha_{\rm sp}$). It is worth mentioning that the sudden variation in the UL gain around $\alpha_{\rm sp}$ proves the significance of the factor $\lambda \sqrt{\frac{P_{\rm d}}{\rho}}$ to the UL rate expression in \eqref{the_only}.

\begin{table} []
\caption{\; Parameters Values.}
\centering
\begin{tabular}{|l|l|l|l|l|l|l|l|}
\hline
\rowcolor[HTML]{C0C0C0}
\textbf{Parameter} & \textbf{Value}      & \textbf{Parameter} & \textbf{Value} & \textbf{Parameter} & \textbf{Value}      & \textbf{Parameter} & \textbf{Value}  \\ \hline
$P^{(\rm{M})}_{\rm{u}}$        &  1 W                & $P_{\rm{d}}$              & 5 W & $\lambda$          & 1 $\text{BSs/Km}^2$ & $\rho$   & -70 dBm                \\ \hline
$B_{\rm u}$              & 1 MHz               & $B_{\rm d}$              & 1 MHz  & $\beta$            & -80 dB              & $N_o$              & -90 dBm          \\ \hline
$\omega_1^{\rm (u)}$, $\omega_1^{\rm (d)}$   & 0.5                 & $\omega_2^{\rm (u)} $, $ \omega_2^{\rm (d)}$  & 1 &$S_{\rm d}(f)$   & Sinc                 & $S_{\rm u}(f)$  & Sinc$^2$                 \\ \hline
\end{tabular}
\label{parameters}
\vspace{-0.8cm}
\end{table}

\section{Simulation and Numerical Results} \label{Results}

This section validates the developed framework and presents selected numerical results that reveal important design guidelines for $\alpha$-duplex cellular systems. The validation part is done via an independent Monte Carlo simulation. Each simulation-run generates two independent PPPs, over a $400 \  \text{Km}^2$, with intensities $\lambda$ and $\lambda_{\rm u}$ for the BSs  and UEs, respectively. Each UE is associated to its nearest BS and the available channel pair is assigned to a randomly selected user within each BS's association area. The transmit power of each user in the UL is set according to the power control discussed in Section~\ref{System Model}. SINR results with the proper pulse-shaping, duplexing, and filtering factors are collected for UEs and BSs located within $4 \  \text{Km}^2$ of the origin to avoid edge effects. For error probability analysis, binary phase shift keying (BPSK) modulation is used for both UL and DL. Hence, according to \cite[Table 6.1]{Wireless2005Goldsmith}, the modulations specific parameters are selected as $\omega_1^{\rm (u)}=\omega_1^{\rm (d)}=0.5$ and $\omega_2^{\rm (u)}=\omega_2^{\rm (d)}=1$. Last but not least, the  SI attenuation factor $\beta$ is selected to be $-80$ dB, which fits within the practical range of [-100,-70] dB given in \cite{Full2015Goyal}. Table~\ref{parameters} summarizes the selected values for the parameters used in the simulation and numerical results.
 
\begin{figure}[t]
\centerline{\includegraphics[width=  2.2in]{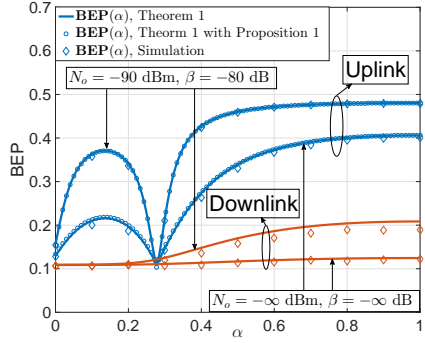}}\caption{\, BEP vs the duplexing parameter $\alpha$ for UL and DL cases.}
\label{fig:BERSim}
 \vspace{-0.8cm}
\end{figure}

Fig. \ref{fig:BERSim} validates the accuracy of the developed analytical model against simulations and captures the BEP behavior for the UL and the DL versus $\alpha$. The figure also confirms the tight bound given by Proposition 1 for the LT of the intra-mode interference for UL CEUs. Fig. \ref{fig:BERSim} manifests the vulnerability of the UL by showing that the effect of the duplexing parameter $\alpha$ is more prominent in the UL scenario.  Assuming perfect SIC (i.e., setting $\beta= -\infty$ dB), while the BEP in the DL is almost unaffected, the UL BEP increased by $216 \%$ at the FD case (i.e., $\alpha=1$). FD transmission with imperfect SIC imposes more significant BEP degradation on both the UL and DL. Hence, FD operation with complete UL/DL overlap may not be the best duplexing scheme due to the cross-mode interference. However, thanks to pulse-shaping, the effect of $\alpha$ is not monotone on the UL BEP.  Fig. \ref{fig:BERSim} shows that there is a range of spectrum overlap that trades BW with  non-significant UL and DL BEP degradation. Interestingly, at  $\alpha=\alpha_{\rm sp}$ the UL/donwlink overlap enhances the BEP in the UL  w.r.t. the HD operation. This is because at $\alpha=\alpha_{\rm sp}$ orthogonality between the UL and DL pulse-shapes is achieved (cf. Figs. \ref{fig:Effective} and \ref{fig:MinAlpha}), which suppresses adjacent channel interference as compared to the HD case.

\begin{figure*}[t!]
    \centering
    \begin{subfigure}[t]{0.31\textwidth}
  \centerline{\includegraphics[width=  2.1in]{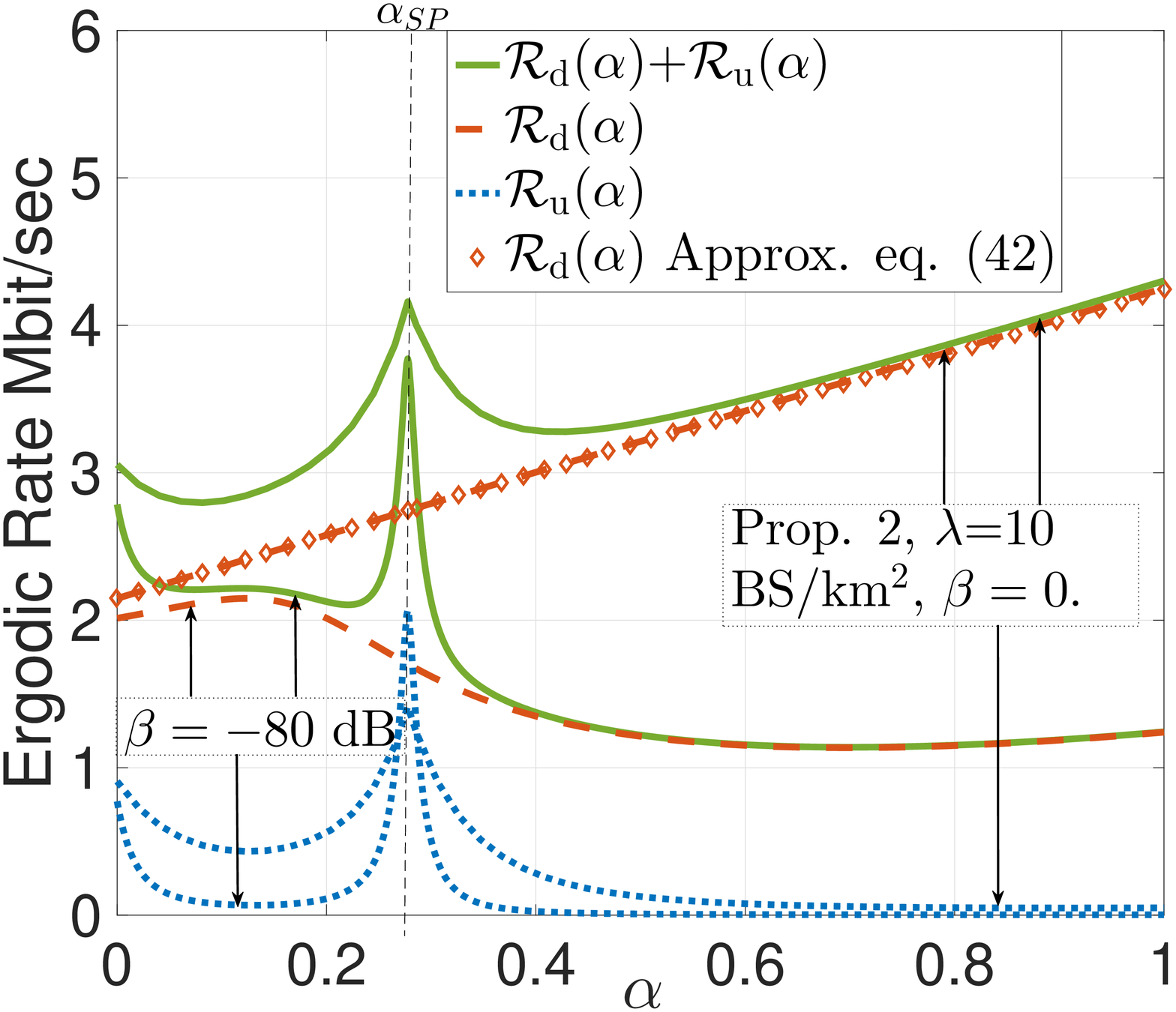}}
      \caption{\,  Ergodic rate for perfect and imperfect SIC.}
\label{fig:Rate1}
    \end{subfigure}%
    ~ 
    \begin{subfigure}[t]{0.31\textwidth}
       \centerline{\includegraphics[width=  2.1in]{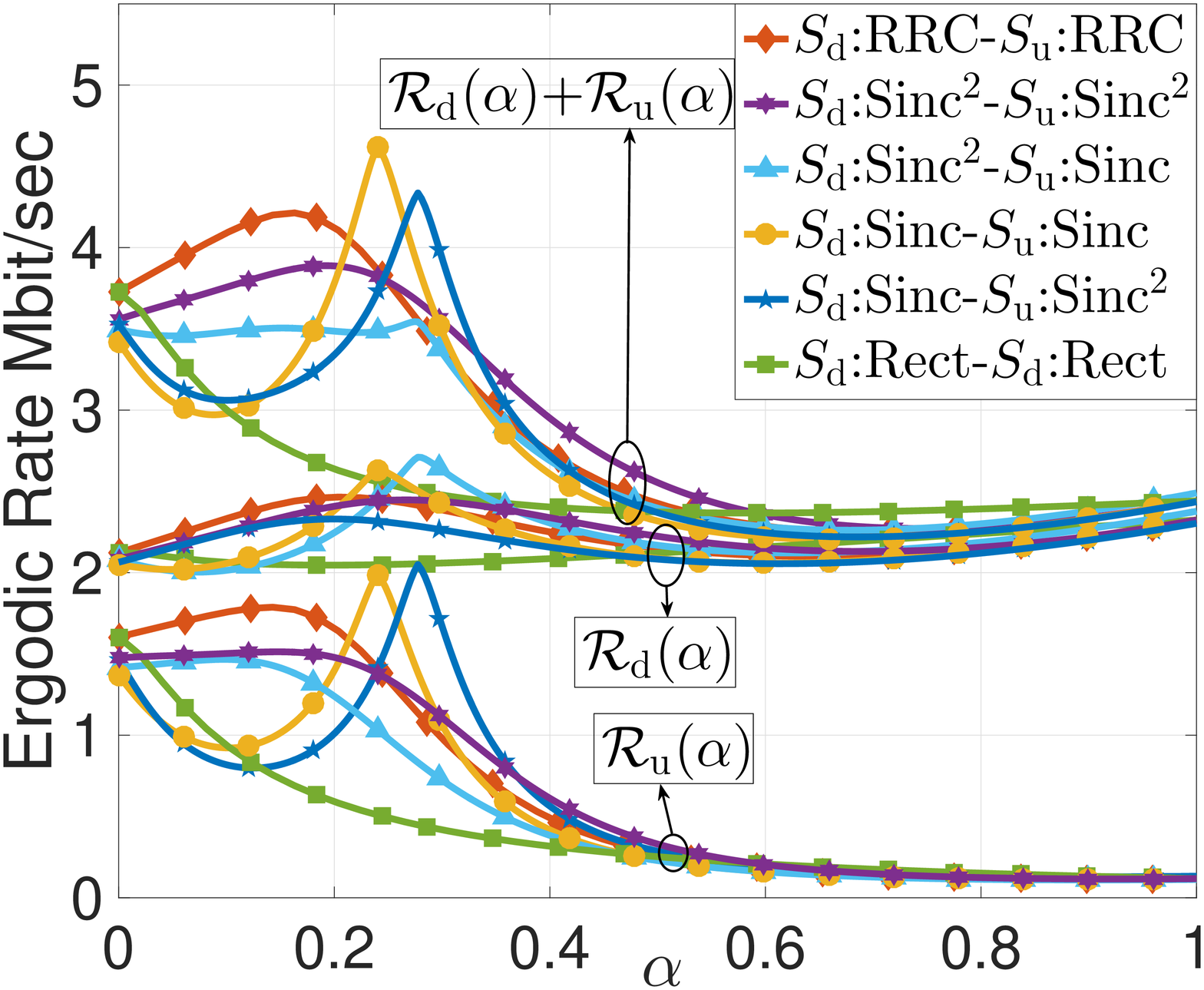}}\caption{\, Ergodic rate vs $\alpha$ for different pulse shapes and $\beta=-95$ dB.}
\label{fig:Pulse_shapes}
    \end{subfigure}
    ~
    \begin{subfigure}[t]{0.31\textwidth}
       \centerline{\includegraphics[width=  2.1in]{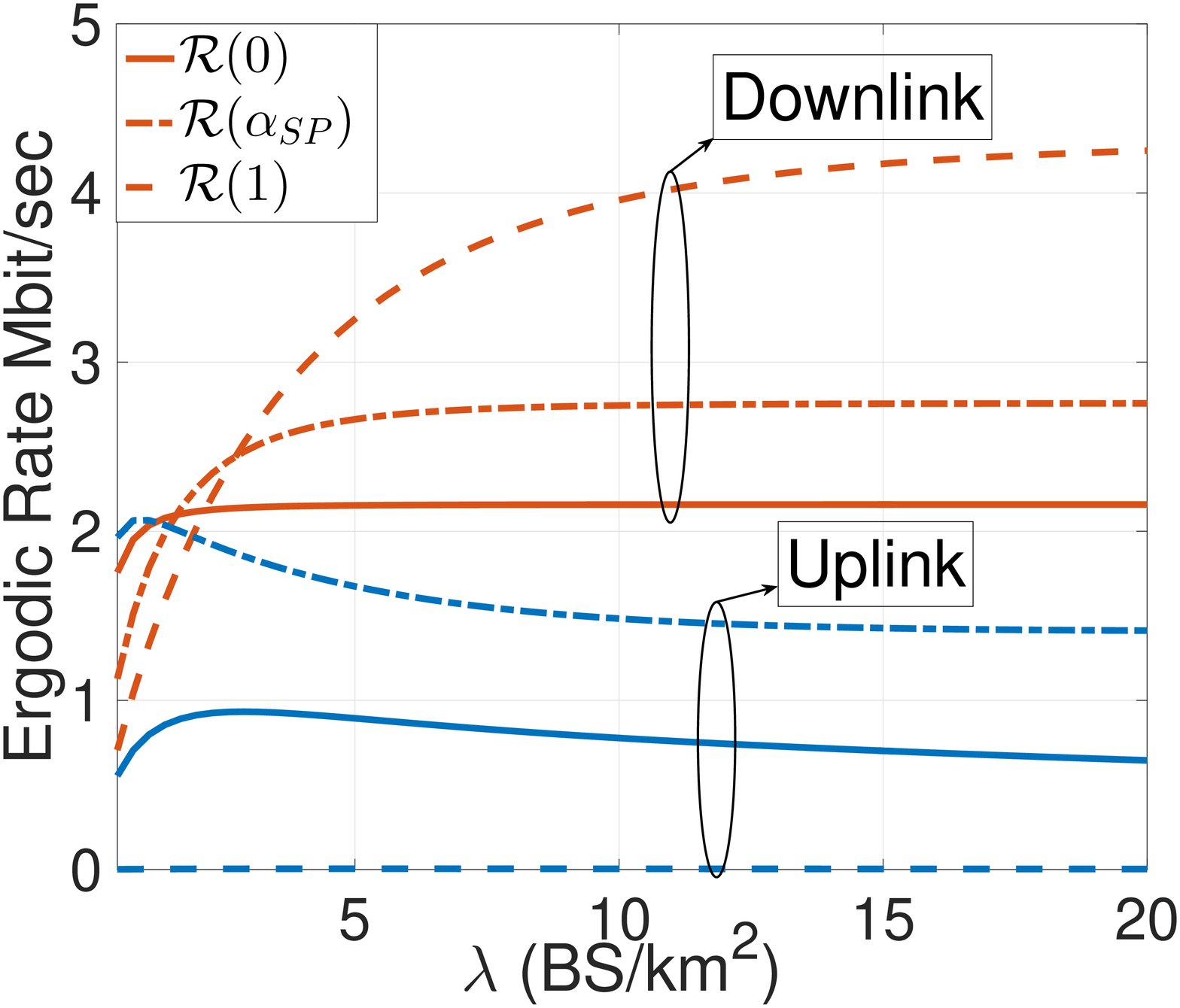}}\caption{\, Ergodic rate vs BSs' intensity in HD, $\alpha_{\rm sp}$, and FD modes.}
\label{fig:Intensity}
    \end{subfigure}
    \caption{The effect of pulse-shaping and matched-filtering on the effective interference factors.}
     \vspace{-0.8cm}
\end{figure*}

Fig.~\ref{fig:Rate1} plots the ergodic rate for the proposed $\alpha$-duplex scheme with Sinc$^2$ and Sinc pulse shapes for the UL and the DL, respectively. The figure shows that both the sum rate and the DL rate improve at FD operation if sufficient SIC is achieved. Hence, overlooking the UL performance leads to misleading  conclusions about the FD operation. Assuming prefect SIC, the figure clearly shows that the performance gain imposed by FD operation in the DL comes at the expense of significant degradation ($94 \%$) in the UL performance. This is because the negative impact of the DL interference on the UL ergodic rate dominates the positive impact obtained by the increased BW. Incorporating the effect of SI, both UL and DL ergodic rates are negatively affected by FD operation. However, due to the UL vulnerability, the degradation in the UL case is much more significant than that of the DL case. The figure also shows that the proposed  $\alpha$ duplexing scheme can improve the spectral efficiency via UL/DL spectrum overlap while alleviating the negative impact of cross-mode interference. The figure also validates the $\arctan(\cdot)$ approximation used in \eqref{the_onlu_Dl1} and \eqref{the_onlu_Dl2} for the DL ergodic rate presented in Proposition 2.  

\normalsize

 In addition to $\alpha$, pulse-shaping is another design parameter for the proposed duplexing scheme. Fig. \ref{fig:Pulse_shapes} shows the UL rate improvement can only be attained at low values of $\alpha$ due to the significance of the cross-mode interference. Furthermore, the ergodic rates achieved by the UL and the DL depend on the used pulse shapes. The figure shows that the Sinc-Sinc$^2$ and the Sinc-Sinc pulse shapes offer the best $\alpha$-duplex UL rate due to the attained orthogonality that nullifies the cross-mode interference on the UL system. The performance of the pulse shapes that nullifies cross-mode interference is followed by the pulse shapes that have the slowest increasing  cross-mode interference factor, namely the RRC-RRC and the Sinc$^2$-Sinc$^2$.  Finally, the Rect-Rect pulse shapes have the fastest increasing cross-mode interference factor, and hence gives the worst $\alpha$-duplex UL rate. Note that the optimal value for $\alpha $ depends on the used pulse shapes, the design objective, and the system parameters as show in Section~\ref{dicsus}. For instance, for perfect SIC and Sinc-Sinc$^2$ pulse shapes, $\alpha=\alpha_{\rm sp}$ is the optimal $\alpha$ for UL maximization objective. However,  $\alpha = 0.25$ is the optimal value of $\alpha$ that maximizes the DL rate (offers $25\%$ improvement) at the cost $0\%$ degradation in the UL rate w.r.t. the HD scheme.


Fig. \ref{fig:Intensity} shows the effect of BSs intensity on the UL and DL rates. The figure shows that FD communication highly degrades the UL rate and that $\alpha$-duplex communication at $\alpha=\alpha_{\rm sp}$ is the best mode for UL operation for all BSs intensities.  On the other hand, the BS intensity determines the best mode of operation for the DL case. Also, the figure shows that the effect of the BSs intensity is monotonic on the DL rate and non-monotonic on the UL rate. In the DL, increasing the BSs intensity increases the intended signal power and increases the DL interference, while it decreases SI\footnote{Due to the power inversion power control, shorter service distance implies less required power for channel inversion.}, and provides a non-monotonic effect on the UL interference. Since the UL to DL interference is negligible, increasing the BS intensity always have a positive effect on the DL rate, specially in the FD mode. This is because the SI, which decreases in the BS intensity, is the performance limiting parameter for the DL FD mode. Note that the intended signal power and the DL interference increases at the same rate \cite{A2011Andrews}, and hence, once the SI becomes negligible w.r.t. the DL interference, the DL rate saturates. Also, the BS intensity controls the relative value of the SI w.r.t. the DL interference, which determines the best mode of operation for the DL.  

The effect of BS intensity shows more complex behavior in the UL direction due to the non-monotonic effect of the BS intensity on both the intended received signal power and the UL interference, which is not negligible in the UL case.  The UL UEs transmit according to the channel inversion power control policy, and hence, varying the BSs intensity varies the transmit power of the CCUs and the received power of CEUs. On the other hand, the received power of the CCU remains constant at $\rho$ and the transmit power of the CEU remains constant at $P^{(\rm{M})}_{\rm u}$. Furthermore, the BS intensity determines the percentage of CEUs and CCUs in the network. In conclusion, when the majority of users are CEUs, the intensity of BSs would have a positive impact on the UL rate, and vice versa. Note that when all UEs become CCUs, the received signal power saturates at $\rho$ and the increased number of interferers is compensated by the decreased transmission power of each and the UL rate saturates.

\subsection{Remarks on the $\alpha$-duplex system design}

 The $\alpha$-duplex scheme relaxes the ``\textbf{all or nothing}'' trade-off, between the BW and cross-mode interference imposed by the FD and HD schemes, to a ``\textbf{fine-tuned}'' trade-off. Furthermore, the proposed scheme enables a flexible cross-mode interference cost assignment for increasing the BW via pulse-shaping and matched filtering. Such adaptable cost function and flexible duplexing are the main reasons for the superiority of the $\alpha$-duplex scheme over both FD and HD schemes. In our future work, we aim to customize the pulse shape to the system parameters and design objective rather than selecting the pulse shape from the set of well known pulse shape templates. This would give more design flexibility and increase the $\alpha$ duplexing gains. Customized pulse shapes can be design following the methodology in \cite{POPS2015Hraiech}.

The results show that the rate gains vs $\alpha$ are different for the UL and the DL according to their sensitivities to the BW and the cross-mode interference as discussed in Section III-F. In particular, the DL rate is more sensitive to the BW than the SINR, and hence, the optimal $\alpha$ is the one that provides the highest BW (i.e., $\alpha =1$). On the other hand,  the UL rate is more sensitive to the SINR than to the BW, and hence, the optimal $\alpha$ is the one that provides the lowest cross-mode interference cost.  Thanks to the pulse-shaping, the results reveals the orthogonality when using the Sinc-Sinc$^2$ pulses, in which the orthogonality is obtained for the system using the Sinc$^2$ pulse. Hence, assigning the Sinc$^2$ to the UL and the Sinc to the DL results in a simultaneous improvement for both systems. In this case, UL rate is improved due to the increased BW at no cross-mode interference cost. Also, the DL rate improves due to its immunity to the UL interference.  

 It is worth mentioning that the optimal $\alpha$ from the sum rate perspective is not necessarily the one that maintains the balance between the UL and DL performances as shown in the results for perfect SI case.   This is because the DL have a higher contribution to the sum-rate and hence, $\alpha=1$ is the optimal value for both the DL and sum rates. However, $\alpha=1$ almost nullifies the UL rate. On the other hand, operating the system at $\alpha = \alpha_{\rm sp}$ balances the trade-off between the UL and the DL rates. Fortunately, the loss in the sum rate at $\alpha = \alpha_{\rm sp}$ is not significant from the $\alpha =1$ case, thanks to the pulse-shaping and the non-negligible contribution of the UL to the sum-rate.

Adding the effect of the imperfect SIC, neither the FD nor the HD operation provides the optimal performance for any of the UL, the DL, or the sum rates. In this case, the optimal value of $\alpha$ can be decided according to the network objective. For SIC of $-80$ dB, Fig.\ref{fig:Rate1} shows that $\alpha = \alpha_{\rm sp}$ improves the UL by  $168 \%$ but degrades the DL by $15 \%$, which results in an overall sum-rate improvement of $36 \%$. Hence, $\alpha = \alpha_{\rm sp}$  is the optimal scheme for UL and/or sum rate maximization objective\footnote{Note that the results obtained in this work coincide with our results in \cite{Can2016ElSawy}, where the results are obtained using simulations for actual BSs deployment in midtown London with practical pathloss values.}.

\section{Conclusion}\label{Conclusion}

In this paper, we develop a tractable framework for in-band FD communications in cellular networks, which explicitly accounts for each of the uplink (UL) and downlink (DL) performances. The developed model is used to shed light on the vulnerability of the UL operation and to show that the cross-mode (i.e., DL to UL) interference is the bottleneck for FD operation in cellular networks. Therefore, we propose a fine tuned duplexing scheme, denoted as $\alpha$-duplex, to maintain a balanced UL and DL operation. The proposed duplexing scheme allows partial overlap between UL and DL while using pulse-shaping and matched filtering to suppress the negative impact of cross-mode interference. Furthermore, we show that the proposed duplexing scheme can help improving self-interference cancellation which further improves the harvested gains. To this end, the results show that neither the traditional half-duplex nor the FD communications provide optimal network operation. Instead, there exists an optimal value for the overlap parameter $\alpha$ which depends on the network parameters and design objective. For instance, we show that cross-mode interference on the UL may be canceled at a certain value for $\alpha$ that achieves orthogonality between UL and DL pulse shapes. At this point, the UL rate is enhanced by $56\%$ and the DL rate is also improved by $28\%$, wherein FD operation, the DL rate in enhanced by up to $97\%$ but at the expense of $94\%$ degradation in the UL rate. We also show that $\alpha$ can be selected to maximize the DL rate subject to a certain degradation constraint in the UL rate. Finally, the effects of pulse-shaping, BS intensity, transmit powers value, and SIC on the network performance are investigated.

\appendices
\section{Proof of lemma 1}
First, we will start by CCUs, note that for CCUs we substitute $P_{r_o}$ by $\rho$. The LT $\mathcal{L}_{\mathcal{I}^{({\rm CCU})}_{\rm u \rightarrow u}}(s)$ can be expressed as, given that the received power from any interferer is less than $\rho$ based on our system model, \cite{Load2014AlAmmouri},
\small
\begin{align}\label{UpIuu1}
 & \mathcal{L}_{\mathcal{I}^{({\rm CCU})}_{\rm u \rightarrow  u}}(s) =   \mathbb{E}_{\tilde{\Psi}_{\rm u}}\left[\underset{x_i \in \tilde{\Psi}_{\rm u}\setminus \{o\}}{\prod} \mathbb{E}_{P_{{\rm u},i},h_i}\left[e^{-s \mathbbm{1}\left( \left\|x_i\right\|> \left(\frac{P_{{\rm u},i}}{\rho}\right)^\frac{1}{\eta} \right) P_{{\rm u},i} h_i \left\|x_i\right\|^{-\eta}}\right]\right]\notag \\
 &\stackrel{(i)}{=}  \exp\left( - 2 \pi  \lambda \int_{\left(\frac{P_{\rm{u}}}{\rho}\right)^\frac{1}{\eta}}^{\infty}\mathbb{E}_{P_{\rm{u}},h}\left[  \left(1- e^{-s P_{\rm{u}} h x^{-\eta}}\right)\right] xdx \right)\notag \\
  &\stackrel{(ii)}{=}\exp \left( - \frac{2 \pi  \lambda}{\eta-2} s \rho^{\frac{-2}{\eta}+1} \mathbb{E}_{P_{\rm{u}}}\left[ P_{\rm{u}}^{\frac{2}{\eta}}\right] {}_2 \text{F}_{1}\left(1,1-\frac{2}{\eta},2-\frac{2}{\eta},-s \rho \right) \right).
\end{align}\normalsize
where $(i)$ follows from the independence between $\tilde{\mathbf{\Psi}}_{\rm u}$ and $h_i$ and using the probability generation functional of PPP. $(ii)$ using the LT of $h$ which is exponentially distributed with unity mean.

For $\mathcal{L}_{\mathcal{I}^{({\rm CCU})}_{\rm d \rightarrow  u}}(s)$ and $\mathcal{L}_{\mathcal{I}^{({\rm CCU})}_{\rm d \rightarrow  u}}(s)$, the interfering BS could be located anywhere around the receiving BS, so there is no interference protection region and since both of them have the same expression, they are equal and can be found as
\small
\begin{align}\label{UpIdu1}
  & \mathcal{L}_{\mathcal{I}^{({\rm CCU})}_{\rm d \rightarrow u}}(s) =\mathcal{L}_{\mathcal{I}^{({\rm CEU})}_{\rm d \rightarrow u}}(s)= \mathbb{E}\left[e^{-  s\underset{x_i \in \tilde{\mathbf{\Psi}}\setminus \{o\}}{\sum}  P_{{\rm{d}},i} h_i \left\|x_i\right\|^{-\eta}}\right]\notag \\
  &\stackrel{(i)}{=}  \exp\left( - 2 \pi  \lambda \int_{0}^{\infty}\mathbb{E}_{h}\left[  \left(1- e^{-s P_{\rm{d}} h x^{-\eta}}\right)\right] xdx \right)\notag \\
  &\stackrel{(ii)}{=}  \exp\left( - \frac{2}{\eta}\pi^{2} \lambda \left( s P_{\rm{d}}  \right)^{\frac{2}{\eta}} \text{csc} \left( \frac{2 \pi}{\eta} \right) \right) .
\end{align}

\normalsize
For $\mathcal{L}_{\mathcal{I}^{({\rm CEU})}_{\rm u \rightarrow u}}(s)$, following \cite{Load2014AlAmmouri}, the interference protection region can be approximated by $||x||<r_o$, so the LT is given by,
\small
\begin{align}\label{eq:UpIuu2}
 \mathcal{L}_{\mathcal{I}^{({\rm CEU})}_{\rm u \rightarrow u}}(s|r_o)  &= \mathbb{E}\left[e^{-  s \underset{x_i \in \mathbf{\Phi}\setminus \{o\}}{\sum} \mathbbm{1}\left( \left\|x_i\right\| > r_o \right) P_{{\rm u},i} h_i \left\|x_i\right\|^{-\eta}}\right]\notag \\
 &\stackrel{(i)}{=}\exp\left( -2 \pi \lambda \int_{r_o}^{\infty}\mathbb{E}_{P_{\rm{u}},h}\left[  \left(1- e^{-s P_{\rm{u}} h x^{-\eta}}\right)\right] xdx \right)\notag \\
 &\stackrel{(ii)}{=}\exp \left(\mathbb{E}_{P_{\rm{u}}}\left[ \frac{-2 \pi \lambda s P_{\rm{u}} r_o^{2-\eta}}{\eta-2}  {}_2 \text{F}_{1}\left(1,1-\frac{2}{\eta},2-\frac{2}{\eta},-s P_{\rm{u}} r_o^{-\eta}\right)\right] \right).
\end{align}
\normalsize

\section{Proof of lemma 2}
For $\mathcal{L}_{\mathcal{I}^{({\rm CCU})}_{\rm d \rightarrow d}}(s)$ and $\mathcal{L}_{\mathcal{I}^{({\rm CEU})}_{\rm d \rightarrow d}}(s)$, the interference protection region is defined by $\left\|x_i\right\|< r_o$ due to the closest BS associations, following the same steps as in Appendix A,

\small
\begin{align}
   \mathcal{L}_{\mathcal{I}^{({\rm CCU})}_{\rm d \rightarrow d}}(s|r0) &=\mathcal{L}_{\mathcal{I}^{({\rm CEU})}_{\rm d \rightarrow d}}(s|r0)= \mathbb{E}\left[e^{- s\underset{x_i \in \mathbf{\Psi} \setminus \{o\}}{\sum} \mathbbm{1}\left( \left\|x_i\right\|> r_o \right)  P_{\rm{d}_i} h_i \left\|x_i\right\|^{-\eta}}\right]\notag \\
&\stackrel{(i)}{=}  \exp\left( - 2 \pi  \lambda \int_{r_o}^{\infty}\mathbb{E}_{h}\left[  \left(1- e^{- s P_{\rm{d}} h x^{-\eta}}\right)\right] xdx \right)\notag \\
 &\stackrel{(ii)}{=}  \exp\left(\frac{-2 \pi \lambda r_o^{2-\eta} P_{\rm{d}} s}{\eta-2} {}_2 \text{F}_1 \left(1,1-\frac{2}{\eta},2-\frac{2}{\eta},-s P_{\rm{d}} r_o^{-\eta} \right)\right).
\end{align}

\normalsize
For $\mathcal{L}_{\mathcal{I}^{({\rm CCU})}_{\rm u \rightarrow d}}(s)$, we approximate the location of the tagged UE to be the same as its serving BSs location (collocated) as in \cite{Hybrid2015Lee}, since the distance between them is limited by $\rm{R_{\rm{M}}}$. Based on this approximation it is given by
\small
\begin{align}
  &\!\!\!\!\!\!\!\!\!\!\!\! \mathcal{L}_{\mathcal{I}^{({\rm CCU})}_{\rm u \rightarrow d}}(s) = \mathbb{E}\left[e^{-  s\underset{x_i \in \mathbf{\Phi}\setminus \{o\}}{\sum} \mathbbm{1}\left( \left\|x_i\right\|> \left(\frac{P_{{\rm u},i}}{\rho}\right)^\frac{1}{\eta} \right) P_{{\rm u},i}   h_i \left\|x_i\right\|^{-\eta}}\right] \stackrel{(i)}{=}  \exp\left( - 2 \pi  \lambda \int_{(\frac{P_{\rm{u}}}{\rho})^{\frac{1}{\eta}}}^{\infty}\mathbb{E}_{P_u,h}\left[  \left(1- e^{- s P_{\rm{u}}   h x^{-\eta}}\right)\right] xdx \right)\notag \\
  &\stackrel{(ii)}{=}  \exp\Bigg( - \frac{2\pi  \lambda s  \mathbb{E}_{P_{\rm{u}}}\left[{P_{\rm{u}}}^{\frac{2}{\eta}} \right] \rho^{1-\frac{2}{\eta}} }{\eta-2 } {}_2 \text{F}_1 \left(1,1-\frac{2}{\eta},2-\frac{2}{\eta},-s \rho \right) \Bigg).
\end{align}

\normalsize
For $\mathcal{L}_{\mathcal{I}^{({\rm CEU})}_{\rm u \rightarrow d}}(s)$, their is no interference protection region so the interferer can be located anywhere, so it is given by

\small
\begin{align}
   \mathcal{L}_{\mathcal{I}^{({\rm CEU})}_{\rm u \rightarrow d}}(s) =&  \exp\left( - 2 \pi  \lambda \int_{0}^{\infty}\mathbb{E}_{P_{\rm{u}},h}\left[  \left(1- e^{-s P_{\rm{u}}   h x^{-\eta}}\right)\right] xdx \right)=  \exp\Bigg( -  \frac{ 2 }{\eta}\pi^{2} \lambda \csc \left(\frac{2 \pi}{\eta} \right) s ^{\frac{2}{\eta}} \mathbb{E}_{P_{\rm{u}}}\left[ P_{\rm{u}}^{\frac{2}{\eta}}\right] \Bigg).
\end{align}

\normalsize
\section{Proof of Proposition 1}

To use Jensen's inequality we have to prove that the argument of the expectation is convex with respect to the RV we are interested in \cite[section 3.1.8]{Convex2004Boyd}. From equation \eqref{LTIuuCEU4}, the expectation is given by,
\small
\begin{align}
\mathbb{E}_{P_{\rm{u}}}\left[\ \sqrt{ P_{\rm{u}} }  \arctan \left(\sqrt{ P_{\rm{u}} } \sqrt{\frac{s  }{r_o^4}} \right)  \right].
\end{align}

\normalsize
Let $y=\sqrt{P_{\rm{u}}}$ be the RV we are interested in and $a=\sqrt{\frac{s  }{r_o^4}}$ a positive constant (with respect to y). Taking the second derivative of the argument results in,
\small
\begin{align}
\frac{d}{dy} \left(y  \arctan \left(a y \right)   \right)=\frac{2 a}{\left(a^2 y^2+1\right)^2}.
\end{align}

\normalsize
Since the second derivative is positive everywhere, then the argument is convex \cite[section 3.1.4]{Convex2004Boyd} and we can use Jensen's inequality which results in,
\small
\begin{align}
\mathbb{E}_{P_{\rm{u}}}\left[\ \sqrt{ P_{\rm{u}} }  \arctan \left(\sqrt{ P_{\rm{u}} } \sqrt{\frac{s  }{r_o^4}} \right)  \right] \geq  \mathbb{E}_{P_{\rm{u}}}\left[\ \sqrt{ P_{\rm{u}} } \right] \arctan \left(\mathbb{E}_{P_{\rm{u}}}\left[\ \sqrt{ P_{\rm{u}} } \right] \sqrt{\frac{s  }{r_o^4}} \right).
\end{align}

\normalsize
Substituting in equation \eqref{LTIuuCEU4} results in equation \eqref{LTIuuCEU4app}.

\section{Proof of Theorem 1}
Since the expressions we found for the $\rm{BEP}$ are in the form of $a \text{erfc}\sqrt{\frac{c x}{y+b}}$ , we can use the following Lemma which is given by \cite{A2009Shobowale} to find the average,
\small
\begin{align}
    \mathbb{E} \left[a \text{erfc}\sqrt{\frac{c x}{y+b}} \right]= a- \frac{a}{\sqrt{\pi} } \int\limits_{0}^{\infty} \frac{\mathcal{L}_y\left(\frac{z}{c}\right) e^{-z(1+\frac{b}{c})}}{\sqrt{z}}dz,
    \label{equ:BE}
\end{align}\normalsize
where $x$ is an exponential RV with unity mean, $y$ a non-negative RV with LT $\mathcal{L}_y(.)$ that is independent of $x$. $b$, $a$ and $c$ are constants. Starting with the first term in eq which is given by,
\small
\begin{align}\label{eq:App1}
\mathbb{E}\left[\omega_1 \text{erfc}\sqrt{\omega_2 \frac{P_{\rm{d}} h_o r_{c}^{-\eta}}{\mathcal{I}^{({\rm CCU})}_{\rm d \rightarrow d} + \mathcal{I}^{({\rm CCU})}_{\rm u \rightarrow d} |\tilde{\mathcal{C}}_{\rm{d}}(\alpha)|^2+ {{\beta} \rho  |\mathcal{\tilde{C}}_{\rm{d}}(\alpha)|^2} + N_o}}\right].
\end{align}

\normalsize
By projecting equation \eqref{eq:App1} on equation \eqref{equ:BE} while conditioning on $r_{\rm c}$, we have $x=h_o$, $y=\mathcal{I}^{({\rm CCU})}_{\rm d \rightarrow d} + \mathcal{I}^{({\rm CCU})}_{\rm u \rightarrow d}$ and $b={{\beta} \rho  |\mathcal{\tilde{C}}_{\rm{d}}(\alpha)|^2} + N_o$. Since in this case $\mathcal{L}_y(s)=\mathcal{L}_{\mathcal{I}^{({\rm CCU})}_{\rm d \rightarrow d}}(s)\mathcal{L}_{\mathcal{I}^{({\rm CCU})}_{\rm u \rightarrow d}}(s)$ then the $\rm{BEP}$ can be easily found by the following equation.
\small
\begin{align}\label{eq:App2}
&\mathbb{E}\left[\omega_1 \text{erfc}\sqrt{\omega_2 \frac{P_{\rm{d}} h_o r_{c}^{-\eta}}{\mathcal{I}^{({\rm CCU})}_{\rm d \rightarrow d} + \mathcal{I}^{({\rm CCU})}_{\rm u \rightarrow d} |\tilde{\mathcal{C}}_{\rm{d}}(\alpha)|^2+ {{\beta} \rho  |\mathcal{\tilde{C}}_{\rm{d}}(\alpha)|^2} + N_o}}\right]= \omega_1^{({\rm{d}})} -  \notag \\
&\int\limits_{0}^{\infty}\frac{ \omega_1^{({\rm{d}})}\mathcal{L}_{\mathcal{I}^{({\rm CCU})}_{\rm d \rightarrow  d}}\left(\frac{z r_{\rm c}^{\eta}}{ P_{\rm{d}} \omega_2^{({\rm{d}})}} \right) \mathcal{L}_{\mathcal{I}^{({\rm CCU})}_{\rm u \rightarrow  d}}\left(\frac{r_{\rm c}^{\eta}z|\tilde{\mathcal{C}}_{\rm{d}}(\alpha)|^2}{P_{\rm{d}} \omega_2^{({\rm{d}})}}\right) } { \sqrt{\pi z} } \exp \Bigg(-z\left(1+ \frac{{\beta} \rho  |\mathcal{\tilde{C}}_{\rm{d}}(\alpha)|^2 r_{\rm c}^{2 \eta}}{\omega_2^{({\rm{d}})} P_{\rm{d}}  } + \frac{N_o r_{\rm c}^{\eta}}{{ \omega_2^{({\rm{d}})}  P_{\rm{d}} }}\right) \Bigg)dz  
\end{align}

\normalsize
Then we have to average over $r_{\rm c}$ by multiplying by its PDF and integrating over its region. By following similar approach, other equations can be obtained.

\section{Proof of Theorem 2}
\normalsize
For the DL case, we know from \eqref{SINR_DL} that the ${\rm SINR}$ is given by
\small
\begin{align}  \label{SINR_DL2}
& {\rm SINR_d} \left( \Xi_{\rm{d}}\right) = \frac{h_o}{\underset{{k \in \Psi}}{\sum} \frac{h_k r_{k}^{-\eta}}{ r_o^{-\eta}} + \underset{j \in \Phi}{\sum} \frac{{P_{{\rm{u}}_j} h_j r_j^{-\eta}} |\mathcal{\tilde{C}}^{({\rm{d}})}_{\rm{u}}(\alpha)|^2}{{P_{\rm{d}} r_o^{-\eta}}}+ \frac{{\beta} P_{{\rm{u}}_o}  |\mathcal{\tilde{C}}^{({\rm{d}})}_{\rm{u}}(\alpha)|^2}{P_{\rm{d}}  r_o^{-\eta}} + \frac{\sigma_n^2}{{P_{\rm{d}} r_o^{-\eta}}}}.
\end{align}

\normalsize
Using the fact that $h_o$ is exponentially distributed with unit mean, the outage probability reduces to
\small
\begin{align}
\!\!\!\!\!\!\!\!\!\!\!\! \mathcal{O}(\theta)&=1-\mathbb{E}_{\Xi_{\rm{d}}}\left[ \exp \left(- \theta \left( \underset{{k \in \Psi}}{\sum} \frac{h_k r_{k}^{-\eta}}{ r_o^{-\eta}} + \underset{j \in \Phi}{\sum} \frac{{P_{{\rm{u}}_j} h_j r_j^{-\eta}} |\mathcal{\tilde{C}}_{\rm{d}}(\alpha)|^2}{{P_{\rm{d}} r_o^{-\eta}}}+ \frac{{\beta} P_{{\rm{u}}_o}  |\mathcal{\tilde{C}}_{\rm{d}}(\alpha)|^2}{P_{\rm{d}}  r_o^{-\eta}} + \frac{\sigma_n^2}{{P_{\rm{d}} r_o^{-\eta}}}\right) \right)\right]\notag \\
&=1-\mathbb{E}_{P_{\rm{u_o}}}\Bigg[ \mathcal{L}_{\mathcal{I}_{\rm d \rightarrow  d}}\left(  \frac{\theta r_o^{\eta}}{P_{\rm{d}}}\right) \mathcal{L}_{\mathcal{I}_{\rm u \rightarrow  d}}\left(\frac{\theta |\mathcal{\tilde{C}}_{\rm{d}}(\alpha)|^2 r_o^{\eta}}{P_{\rm{d}}}\right) \Bigg(- \theta \left(\frac{{\beta} P_{\rm{u_o}}  |\mathcal{\tilde{C}}_{\rm{d}}(\alpha)|^2 r_o^{ \eta}}{ P_{\rm{d}}  } + \frac{\sigma_{n}^2 r_o^{\eta}}{{  P_{\rm{d}} }}\right) \Bigg) \Bigg].
\end{align}

\normalsize
By substituting $P_{\rm{u_o}}$ by $\rho$ for CCUs and by $P_{\rm{u}}^{(\rm{M})} r_o^{-\eta}$ for CEU, equations in Theorem 2 can be found for the DL. For the UL case, similar approach can be followed to obtain the expressions.

\bibliographystyle{IEEEtran}
\bibliography{ref}

\vfill

\end{document}